\documentclass[a4paper]{article}

\usepackage[utf8]{inputenc}
\usepackage[T1]{fontenc}
\usepackage[english]{babel}

\usepackage{amsmath}
\usepackage{amssymb}
\usepackage{mathrsfs}

\usepackage{graphicx}

\usepackage{url}

\usepackage{afterpage}

\usepackage{subfig}
\usepackage{float}
\usepackage{authblk}

\usepackage{color}
\usepackage{fixltx2e}
\usepackage{textcomp}
\usepackage{cite}
\usepackage[normalem]{ulem}

\usepackage{siunitx}
\DeclareSIUnit\torr{torr}

\begin{document}
\newcommand{\fref}[1]{Fig.~\ref{#1}}
\newcommand{\new}[1]{\textcolor{green}{#1}}

\title{Non-Ising domain walls in \textit{c}-phase ferroelectric lead titanate thin films}

\author[1]{Weymann Christian}
\author[2]{Cherifi-Hertel Salia}
\author[1]{Lichtensteiger Céline}
\author[1]{Gaponenko Iaroslav}
\author[2]{Dorkenoo Kokou Dodzi}
\author[3]{Naden Aaron B.}
\author[1]{Paruch Patrycja}
\affil[1]{Departement of Quantum Matter Physics, University of Geneva, 24 Quai Ernest-Ansermet, CH-1211 Geneva 4, Switzerland}
\affil[2]{Université de Strasbourg, CNRS, Institut de Physique et Chimie des Matériaux de Strasbourg, UMR 7504, Strasbourg, 67000, France}
\affil[3]{School of Chemistry, University of St. Andrews, United Kingdom}

\date{\today}

\maketitle

\begin{abstract}
Ferroelectrics are technologically important, with wide application in micromechanical systems, nonlinear optics, and information storage. Recent discoveries of exotic polarisation textures in these materials, which can strongly influence their properties, have brought to the forefront questions about the nature of their domain walls -- long believed to be primarily Ising, with locally null polarisation. Here, combining three complementary techniques -- second harmonic generation microscopy, piezoresponse force microscopy, and transmission electron microscopy - to cover all the relevant lengthscales, we reveal the Néel character (non-Ising polarisation oriented perpendicular to the wall) of \SI{180}{\degree} domain walls in \textit{c}-phase tetragonal ferroelectric lead titanate epitaxial thin films, for both artificial and intrinsic domains at room temperature.  Furthermore, we show that variations in the domain density -- detected both optically and via local piezoresponse, then quantified by radial autocorrelation analysis -- can give us insight into the underlying defect potential present in these materials.

\end{abstract}

Ferroelectric materials are characterised by their spontaneous and reversible electrical polarisation, with a wide variety of technological applications as microelectromechanical systems (MEMS) and memory devices \cite{Setter-JoAP-2006}. The electromechanical properties of these materials are closely linked to their domain structure. As such, understanding and controlling the arrangement of domains with different polarisation orientation remains a key research objective. In ultrathin films, where strong depolarising fields destabilise the polarisation and drive the formation of ever smaller domains with increasing densities of domain walls, strongly modified dielectric and piezoelectric responses \cite{Zhang-JAP-1994, Damjanovic-1998, Sluka-NC-2013} as well as effective negative capacitance \cite{Zubko-N-2016, Lukyanchuk-PRB-2018} have been observed. In addition, the domain walls themselves can present emergent properties \cite{Seidel-NatureMat-2009,Guyonnet-AdvMat-2011,Daraktchiev-PRB-2010} potentially attractive for domain-wall-based nanoelectronics \cite{Catalan-RMP-2012,Meier-NM-2012}. Their complex (de)pinning dynamics when driven by an electric field \cite{Paruch-CRP-2013} have in particular given rise to varied implementations as nanoscale active device components \cite{Whyte-AM-2014,McGilly-NN-2015}.

While ferroelectric materials were long believed -- in contrast to their ferromagnetic counterparts -- to present essentially Ising type \SI{180}{\degree} domain walls, in which the polarisation is maintained along the bulk axis, decreasing to zero and reversing its orientation at the domain wall centre, this is now recognised not to be the case. Extensive theoretical studies \cite{Lee-PRB-2009, Marton-PRB-2010, Stepkova-JPCM-2012, Marton-PT-2013, Wang-JoAP-2017, Wojdel-PRL-2014, Eliseev-PRB-2013} predicted non-Ising polarisation components oriented either perpendicular to the wall (Néel type) or in the plane of the wall (Bloch type). Scanning transmission electron microscopy and second harmonic generation measurement at room temperature demonstrated the existence of Néel-type domain walls in Pb(Zr,Ti)O$_3$ \cite{Wei-NatCom-2016, Luca-AM-2017, Cherifi-Hertel-NC-2017}, and of chiral Bloch walls and Bloch lines, a particular type of ferroelectric topological structures, in LiTaO$_3$ \cite{Cherifi-Hertel-NC-2017}. Additional exotic polarisation structures, such as arrays of polar vortices \cite{Yadav-N-2016}, `supercrystals' of highly ordered flux closure domains \cite{Stoica-NM-2019,Hadjimichael--2019,Hadjimichael-NatMat-2021}, and even ferroelectric skyrmions \cite{Goncalves-SA-2019,Das-N-2019} have also recently been reported. In the canonical tetragonal ferroelectric PbTiO$_3$, hosting many of these emergent exotic polarisation textures, various theoretical calculations predict either Néel walls \cite{Lee-PRB-2009,Behera-JPCM-2011} like those observed in Pb(Zr$_{0.2}$Ti$_{0.8}$)O$_{3}$ \cite{Luca-AM-2017,Cherifi-Hertel-NC-2017}, of which it is one of the parent compounds, or a low temperature Ising-to-Bloch transition at the domains walls \cite{Wojdel-PRL-2014}.  Metastable bubble domains observed in PbTiO$_3$ thin films \cite{Zhang-AM-2017}, with no strong delineation between the domain wall and interior, further suggest that the polarisation in this material can be extremely `soft' and easy to reorient in either Bloch- or Néel-like textures under specific electrostatic and strain boundary conditions.

In this context, we set out to characterise the nature of \SI{180}{\degree} domain walls in PbTiO$_3$. Our work is based on three 50 nm PbTiO$_3$ thin films deposited by rf-magnetron sputtering on LaNiO$_3$ electrodes on single crystal (001) SrTiO$_3$ substrates, with an intrinsic domain configuration governed by the interplay between mechanical and electrostatic boundary conditions \cite{NanoLetters-2014, NJP-2016} and additionally controlled via growth temperature defect engineering \cite{Weymann-AEM-2020}. X-ray diffraction measurements, topography and PFM measurements show that all these samples are purely in the $c$-phase -- two samples being in a polydomain state with regions of up- and down-oriented polarisation, and the other in a monodomain up-oriented state. To characterise these samples, we used three complementary techniques: second harmonic generation (SHG), piezoresponse force microscopy (PFM), and transmission electron microscopy (TEM), which taken together allow us to map the polarisation at all the relevant lengthscales. All measurements were carried at room temperature. Using PFM and SHG, we find that Néel-like behaviour dominates at \emph{artificial} domain walls introduced by local electric field switching of polarisation. By combining PFM and careful analysis of TEM data, we show that the same is true for \emph{naturally occurring} domain walls in polydomain samples. Finally, we also show that it is possible to extract information on the underlying pinning potential by additional SHG and PFM image treatment and analysis based on auto-correlation.


\begin{figure}[htbp]
	\centering
	\includegraphics[width=0.8\textwidth]{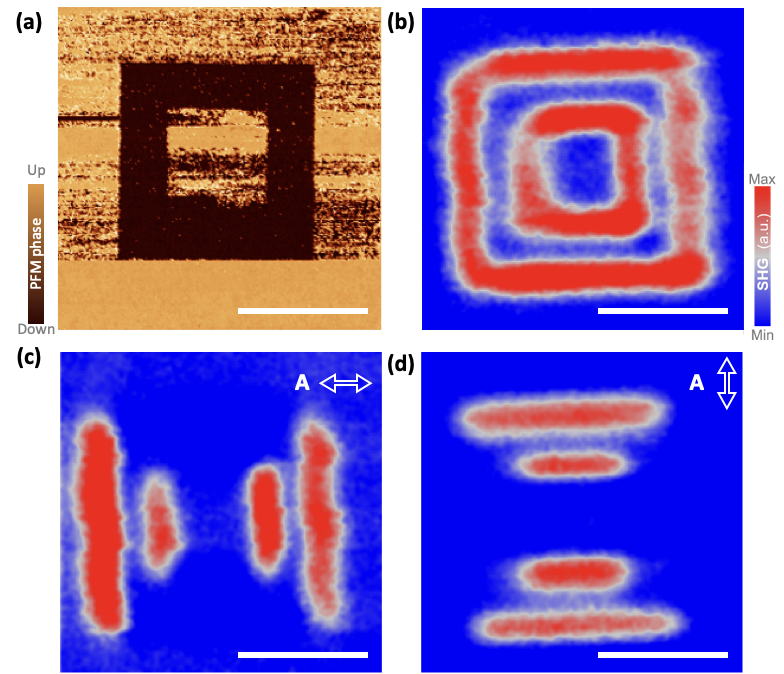}
	\caption{\emph{Signature of Néel polarisation at written domain walls at room temperature} (a) Vertical PFM phase image of the square-in-square domain structure written in a up-oriented monodomain film. (b) Isotropic SHG image (without polarisation analysis) acquired at the patterned square-in-square domain structure, with red corresponding to maximum signal intensity and blue to minimum signal intensity. We observe a localized second-harmonic emission at the domain wall regions, which demonstrates their non-Ising polar character. (c,d) SHG measurements of the domain structure for different polariser and analyser angle combinations, where the analyser direction is represented by the white arrow. SHG signal is observed at the domain walls only when the analyser is perpendicular to the domain wall plane. This polarimetry analysis shows that the local emission is polarised in the direction of the white arrows. This demonstrates that the local polarisation dipole is perpendicular to the domain walls, consistent with a Néel character. The white scale bar in all panels represents \SI{4}{\micro\metre}.}
	\label{Written}
\end{figure}

In the out-of-plane up-oriented monodomain film, domains were patterned by applying relatively high bias to the scanning probe microscopy tip, first $+10$ V to switch a $6 \times 6$ \si{\micro\meter^2} square, then $-10$ V in a $3 \times 3$ \si{\micro\meter^2} square in the center, as shown in \fref{Written}(a), ensuring a well-spaced square-in-square domain structure whose walls could be easily individually resolved by SHG. SHG measurements were then performed to identify the presence and orientation of any in-plane component of the ferroelectric polarisation at the domain walls. As detailed in Ref.~\cite{Cherifi-Hertel-NC-2017,Cherifi-Hertel-JAP-2021}, no SHG signal is expected in the geometry used for this experiment (see Methods) for tetragonal materials (\textit{4mm} point group symmetry). From the anisotropic SHG image (\fref{Written}(b)), a localized second-harmonic emission is observed at the domain wall regions, demonstrating their non-Ising polar character. Additional polarimetry analysis was performed on this region for different polariser and analyser angle combinations. The written domain walls are clearly visible in the SHG image only when the analyser is perpendicular to the wall plane, as shown in \fref{Written}(c-d). We can thus conclude that in this case the domain walls present an in-plane polarization component with a purely Néel-type configuration, as similarly observed in the related material Pb(Zr$_{0.2}$Ti$_{0.8})$O$_3$ \cite{Cherifi-Hertel-NC-2017}. 


\begin{figure}[htbp]
	\centering
	\includegraphics[width=0.55\textwidth]{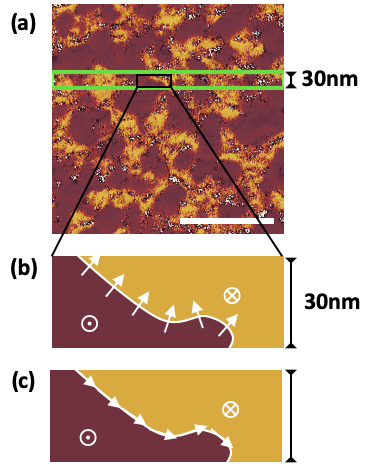}
	\caption{\emph{Top view of nanoscale domains and domain walls} (a) Vertical PFM phase image of the domain structure of a polydomain sample.  The white scalebar represents \SI{400}{\nano\meter}. The typical thickness of the TEM lamella prepared from this sample (\SI{\approx 30}{\nano\meter}) is represented by the green lines. (b-c) Schematic representation of the polarisation in the black rectangle in (a), for the case of Néel (b) or Bloch type (c) domain walls respectively. From this schematic representation, it is clear that domain walls are highly unlikely to align through the TEM lamella thickness, but will rather meander, leading to averaging over different polarisation orientation in the TEM cross-section measurements. For a wall that is not imaged edge-on, ie orthogonal to the plane of the lamella, the two possible polarisation configurations at the wall (b) and (c) could lead to similar projections.}
	\label{Schematic}
\end{figure}

For polydomain samples, vertical PFM imaging shows alternating regions with up- and down-oriented domains, of the order of a few 10s of nm across, as can be seen in \fref{Schematic}(a). The high density and small size of these intrinsic domains make it impossible to individually resolve domain walls and determine their nature via SHG polarimetry, but non-negligeable contrasts in the measured intensity point to the presence of in-plane polarisation (as we will discuss further later). We therefore turn to scanning transmission electron microscopy (STEM) to visualise the individual atomic displacements contributing to the macroscopic polarisation. TEM measurements were performed on a \SI{\approx 30}{\nano\metre} lamella prepared with the long axis parallel to the [100] in-plane axis of the thin film, as schematically indicated by the green box in \fref{Schematic}(a). Capturing a domain wall perfectly edge-on in such a lamella is unlikely, and we thus expect to observe a projection of the wall, in which case the TEM measurements would average over the two opposing out-of-plane polarisation orientations of the domains as well as any Néel or Bloch domain wall polarisation, as schematically indicated in \fref{Schematic}(b) and (c), respectively. Nevertheless, within these constraints we can preselect regions where the edge-on domain wall alignment in the lamella is at least somewhat enhanced.

\begin{figure}[htbp]
	\centering
	\includegraphics[width=0.89\textwidth]{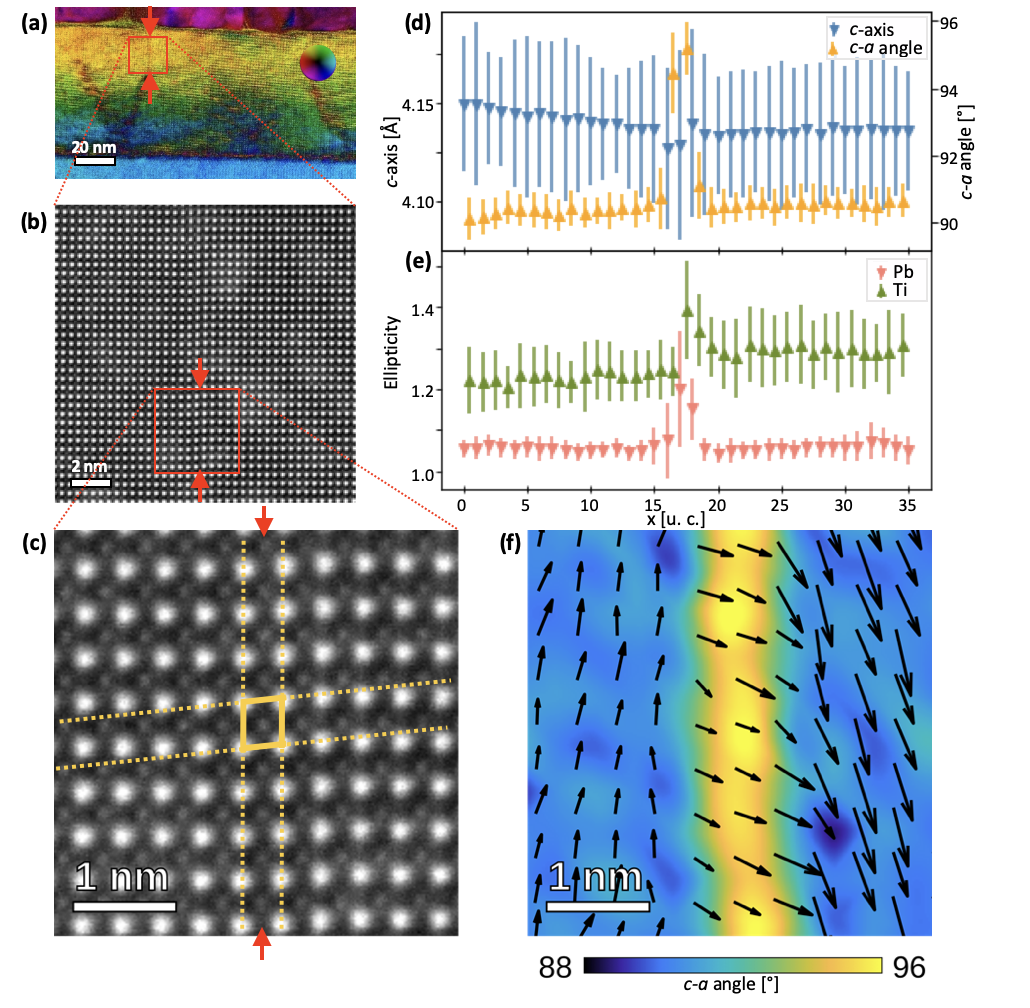}
	\caption{\emph{Transmission electron microscopy and cross-section polarisation mapping of intrinsic domain walls at room temperature} (a) Low-magnification differential phase contrast (DPC) TEM image of the sample, allowing visualisation of the domain walls positions. We choose to focus on the area where the domain wall appears thinnest (red box), since this is where it will be most edge-on. (b) High magnification HAADF-STEM image of the region indicated by a red box in (a). The domain wall is visible as a shift in the Pb lattice close to the center of the image (see main text). (c) Cropped version of the atomic resolution HAADF shown in (b). Here the Ti atom columns are  readily visible, as well as the shift in the Pb lattice, inducing a sharp change in the $c-a$ angle highlighted by the yellow dotted lines. The domain wall position is indicated by red arrows in (a-c). (d) $c$-axis and $c-a$ angle extracted from 2D Gaussian fits to the positions of the atomic columns in (b). The position of the domain wall is clearly visible as an increase in $c-a$ angle. Concomitant with this increase, one can note a drop in local $c$-axis (although within the error bars). (e) Ellipticities of the Pb and Ti columns, caused by the projection of different positions of the atoms throughout the thickness of the lamella. While both show an increase at the position of the wall, the Ti ellipticity is higher throughout, consistent with the averaging over opposite polarisation directions through the thickness of the lamella. (f) Reconstructed polarisation map based on a three Gaussian fit to the position of the Ti columns in (c) (see main text). The color scale indicates the interpolated $c-a$ angle, which we use as a marker for the position of the domain wall. The arrows represent the polarisation extracted from the Ti displacement, which follows a typical Néel pattern: the polarisation rotates through \SI{180}{\degree} between the two domains on either side, such that it is perpendicular to the wall plane at the position of the wall.}
	\label{Intrinsic}
\end{figure}

\fref{Intrinsic}(a) shows a low magnification differential phase contrast (DPC) STEM image where the colour scale indicates the direction and relative magnitude of the deflection of the STEM probe due to diffraction and Lorentz deflection. Within the PbTiO$_3$ film, elongated linear features between regions of relatively homogeneous contrast can be identified, appearing dark red and thus indicating an increased in-plane deflection component. These features appear narrower and vertically oriented near the film surface, then gradually broaden and curve towards the interface of the film with the back electrode and substrate. We therefore focused on the near-surface region, to enhance the probability of their edge-on alignement. The red box demarcates an area imaged at atomic resolution in the high angle annular dark field (HAADF) image in \fref{Intrinsic}(b). Here, the abrupt feature in the centre can be seen to manifest as a distinct shift in the Pb atom positions, resulting in apparent `kink' in the Pb lattice (brightest atom columns). This confirms the prediction of Aguado-Puente and Junquera for \SI{180}{\degree} domain walls in PbTiO$_3$ thin films \cite{Aguado-Puente2012}, and is very similar to the Bi-lattice `kink' observed in BiFeO$_3$ related to the polarisation rotation and accumulation of charged defects at domains walls \cite{Rojac-NatMat-2017}. Furthermore, it can be seen from \fref{Intrinsic}(b) and its cropped version shown in \fref{Intrinsic}(c) that the Ti atom columns sit closer to the centre of the unit cell on the left whereas they are near the bottom of the unit cell on the right. In other words, there is evidently a greater out-of-plane component of polarisation on the right since this is closely linked to the B-site cation displacement from the centrosymmetric position within the unit cell \cite{Shirane-PRB-1970}. This observation therefore suggests the presence of a domain wall in the centre of the image.

In order to accurately determine the atom positions and hence polarisation, we perform 2D Gaussian fitting to the Pb and Ti atom columns which offers subpixel precision \cite{Yankovich-NC-2014, Wang-AdvMaterInterfaces-2015,Bencan-NatComm-2021}. From this, we can determine a number of factors that can confirm the presence of a domain wall: $c$-axis and $c-a$ angle (angle between the [100] and [001] directions), as shown in the averaged line profiles in \fref{Intrinsic}(d) which go from left to right in \fref{Intrinsic}(b). In addition, we can also determine the ellipticity of the atom columns -- shown in \fref{Intrinsic}(e) -- which gives a measure of the disorder through the thickness of the lamella: an ideally ordered column of atoms would appear perfectly circular. From the profiles in \fref{Intrinsic}(d) and (e), it is clear the kink observed in \fref{Intrinsic}(b) and (c) coincides with an increase in the $c-a$ angle and in both the Pb and Ti ellipticities. One can also note a concomitant decrease in out-of-plane lattice parameter $c$ - although within the error bars. Crucially, all of this occurs over approximately 2--3 unit cells, and coincides with an increase in the in-plane component of the polarisation defined by the displacement of the Ti columns from the centre of the unit cell given by the Pb atoms, as shown in \fref{Intrinsic}(e). 

To properly interpret the out-of-plane component of the polarisation, careful analysis has to be performed, bearing in mind the likely scenario of domain overlap through the thickness of the lamella, as depicted in \fref{Schematic}. Indeed, \fref{Intrinsic}(e) shows that the Ti atom columns are significantly more elliptical than the Pb atom columns, indicating a notable variation in the Ti position through the lamella thickness, meaning that the observed Ti position is actually the mean \emph{projected} position. This results in the initial fitting shown in Fig.~S1(c), where rather than the \SI{180}{\degree} polarisation rotation that would be expected on opposing sides of a domain wall, the polarisation goes from strongly `down' on the right to very weakly `down' on the left. 

In order to deconvolute the different domain contributions, we fit three 2D Gaussians to each of the Ti atom columns: two corresponding to Ti atom shifts up and down relative to the mean position given by the single Gaussian fitting and one corresponding to the `mean' position between the first two. This approach is analogous to the `mpfit' algorithm proposed previously \cite{Mukherjee-ASaCI-2020}. We can then reconstruct the dominant polarisation from these three components, discarding extraneous effects from projection.  We note, however, that the displacement of the Ti atoms only corresponds exactly to the polarisation of the PbTiO$_3$ if the Born effective charge is constant, as it would be expected for ferroelectric distortions along the \textit{c}-axis orientation, with the material in the \textit{P4mm} phase in the bulk of the domain, but which may not be the case at the domain wall. For the sake of brevity, a comprehensive discussion of the procedure is detailed in the Supporting Information (Section S1). \fref{Intrinsic}(f) shows the final result of this reconstruction where the image is an interpolated map of the $c-a$ angle with the polarisation vectors overlaid.  Moving from left to right on the image, a classic Néel-like behaviour is observed: the polarisation rotates through \SI{180}{\degree} from upwards on the left to downwards on the right with a predominantly in-plane orientation in the 2--3 unit cells at the Pb `kink' in the centre where the $c-a$ angle is maximised and the tetragonality is minimised.


It therefore appears that in our PbTiO$_3$ thin films, at room temperature, both written and intrinsic domain walls present a Néel character. For the written domain walls, the microscopic mechanisms underlying this configuration may be related to the localised high electric field under the tip, necessary to switch the polarisation, but which can also alter the samples electrochemically \cite{Kalinin-AN-2011} and stabilise the polar discontinuity inherent in the Néel structure. For intrinsic domain walls, however, first-principles calculations predict a Bloch character at low temperature and, at higher temperatures, an Ising character with no in-plane component of polarisation \cite{Wojdel-PRL-2014}. Our observations of a Néel component suggest that in real materials such structures are on the contrary at the very least metastable. The polarisation discontinuity at domain walls with such a Néel component can be related to effects of domain wall curvature or inclination~\cite{Acevedo-Salas-arXiv-2022}, in particular at domain walls reconfigured under high local electric fields. Moreover, even at nominal straight intrinsic domain walls, distinct few-unit-cell steps with head-to-head or tail-to-tail polarisation have been observed~\cite{Jia-Science-2011}. Such structures, with their associated charge, could be potentially important for domain-wall-based nanoelectronics, as they could contribute to a higher conductivity at the domain walls~\cite{Morozovska-Ferro-2012,Eliseev-PRB-2012-85-045312,Feigl-NatCom-2014,Stolichnov-NanoLetters-2015,Cao-APL-2017}.  Moreover, in thin films, which generally present a relatively high density of oxygen vacancies and other internal charged species with respect to single crystals, such discontinuities in the polarisation could very easily be stabilised by these defects.  

Indeed, defects and disorder are a crucial element of ferroelectric thin films, acting as nucleation sites during switching \cite{Jesse-NM-2008}, pinning domain walls \cite{Paruch-CRP-2013,Blaser-APL-2012,Tueckmantel-PRL-2021}, and segregating along them to provide a path for electrical conduction \cite{Farokhipoor-PRL-2011, Guyonnet-AdvMat-2011,Seidel-PRL-2010}. In order to better understand and exploit the functional properties of these ferroelectric domain walls, it is therefore crucial to consider them within the defect landscape they inhabit within the PbTiO$_3$ sample. It was previously shown that variations in Pb and O stoichiometry, tuned by the temperature and electrostatic boundary conditions during growth, can lead to strong asymmetry between the two polarisation orientations and variations in the magnitude of polarisation through the film thicknesses  \cite{Weymann-AEM-2020,Strkalj-NatComm-2020}. In the present study, as can be seen in \fref{Intrinsic}(a), the up-oriented domain appear to increase in volume near the surface of the film, at the expense of the down-oriented domain, with a marked curvature of the domain walls suggesting components of even stronger head-to-head and tail-to-tail polar discontinuities.

\begin{figure}[htbp]
	\centering
	\includegraphics[width=0.7\textwidth]{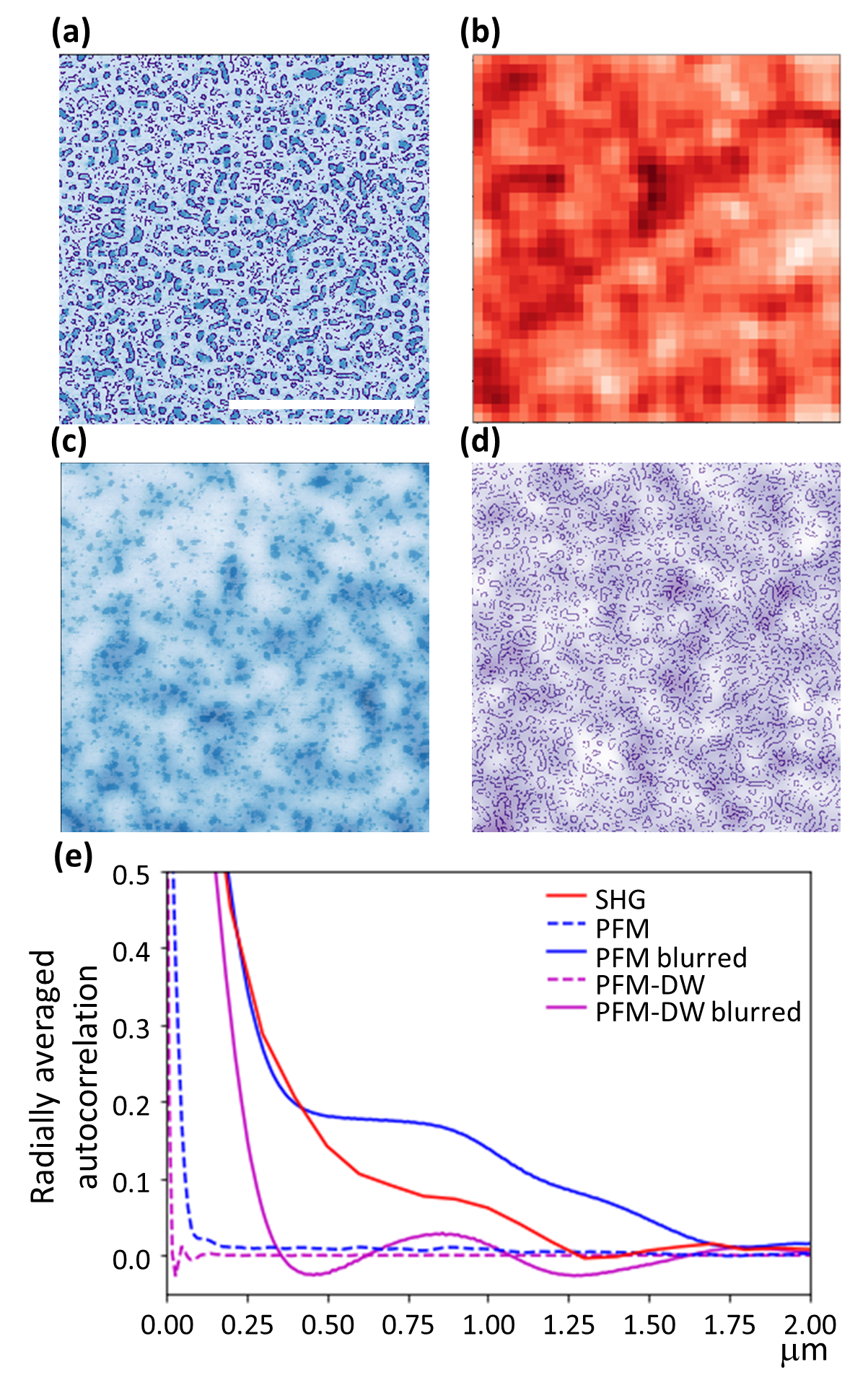}
	\caption{\emph{Characteristic lengthscale of domain superstructure} (a) Vertical PFM phase image of the domain structure of the polydomain sample in blue (ranging from 0 to \SI{180}{\degree}), with the edges detected using the Canny edge detection algorithm in purple. (b) SHG image of the same sample. The dark and bright contrast cannot be directly related to the domain pattern or to the domain edges. (c,d) Local density variations revealed by blurring applied on the vertical PFM phase image of the domains in (c) and on the domain domain wall structure (d).  These domain density variations appear to closely match the variation observed in SHG intensity. All images are shown at the same scale, indicated by the \SI{2}{\micro\metre} white scale bar in (a). (e) Radially averaged autocorrelation for the SHG image and both the domain and domain wall distributions obtained from the PFM phase image, and their superstructure obtained after blurring. While only very small structures are revealed in the raw PFM images, the superstructure in the density has the same pseudoperiod as the contrast variations found in the SHG image.}
	\label{SHGpoly}
\end{figure}

Further insight into the defect landscape can be obtained from a comparative analysis of PFM and SHG measurements in another polydomain sample (\fref{SHGpoly}). The vertical PFM phase response in \fref{SHGpoly}(a) reveals intrinsic domains of the order of \SI{100}{\nano\metre}, well below the diffraction limit for the wavelength used in SHG measurements ($\sim$ \SI{300}{nm}). When comparing with the image obtained by SHG on the same sample and at the same scale (\fref{SHGpoly}(b)), we nonetheless observe regions of brighter and darker SHG contrast. This signal confirms the presence of an in-plane component of polarisation. However, these contrast variations appear at much larger lengthscales than the individual nanoscale domains imaged by PFM. Indeed, when we extract the radially averaged autocorrelation for each image, we find a characteristic pseudo-period of approximately \SI{800}{nm} for the SHG contrast variations, but only about \SI{100}{\nano\metre} for the PFM images, corresponding directly to the domain structure (see Supporting Information Section S2 for details).

Nonetheless, when the microstructure of domains (\fref{SHGpoly}(c)) and domain walls (\fref{SHGpoly}(d)) in the  PFM image is explicitly blurred to mimic the lower resolution and averaging effect of SHG, longer range variations in density are clearly revealed. Radially averaged autocorrelation analysis of this resulting superstructure shows the same approximately \SI{800}{nm} pseudo-period as that of the intensity variations in the SHG image \fref{SHGpoly}(e). 

\begin{figure}[htbp]
	\centering
	\includegraphics[width=\textwidth]{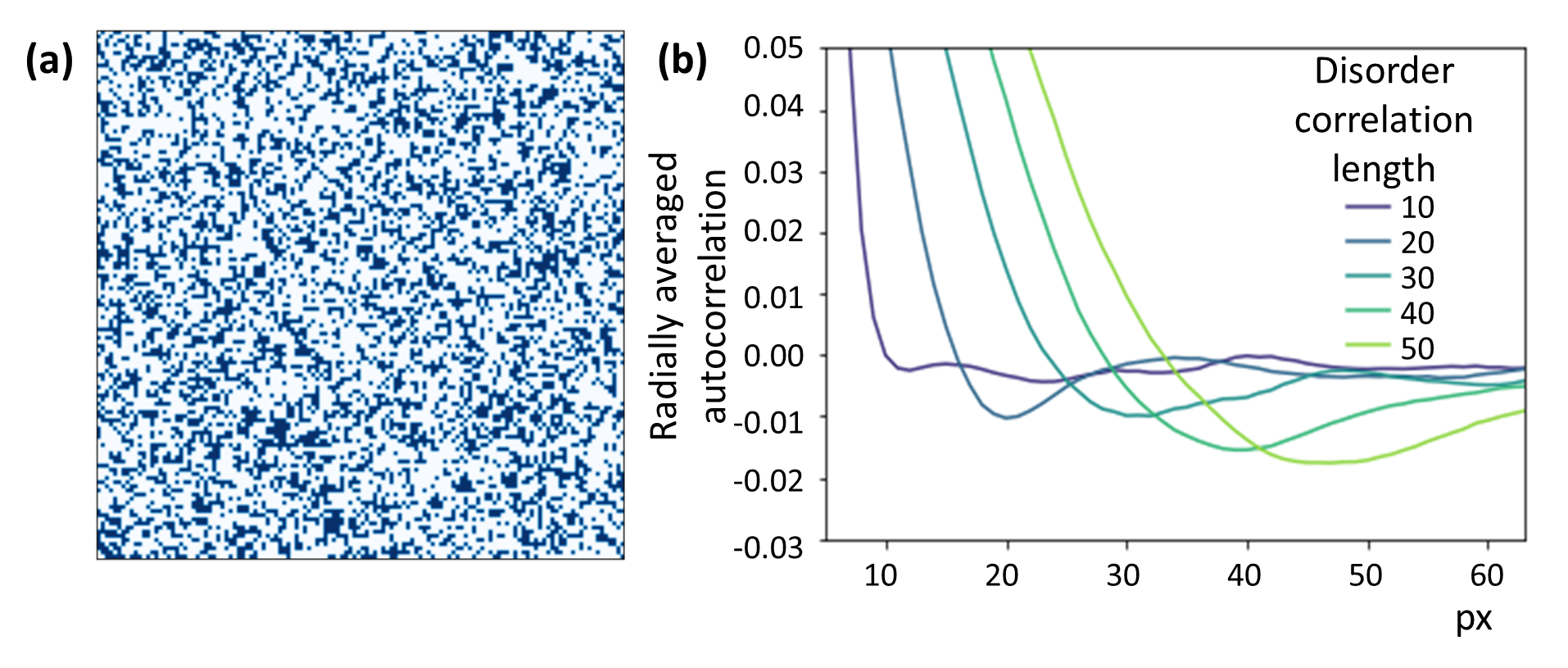}
	\caption{\emph{Simulations of domain density variations demonstrating the importance of the disorder correlation length.} Experimentally observed fluctuations of domain density were reproduced using an Ising system with a random disorder potential of characteristic lengthscale $\xi$. (a) Relaxed domain structure after randomly distributed up and down polarisation regions, initially in equal proportion, were subjected to a random field disorder potential. (b) The radially averaged autocorrelation function of the domain superstructure, whose first minima, corresponding to the typical feature size, exactly match the disorder correlation length.}
	\label{Simulation}
\end{figure}

This superstructure in the domain density appears to be directly linked to the underlying defect landscape. To demonstrate this, we are applying this radially averaged autocorrelation analysis to a simple Ising model simulation shown in \fref{Simulation}(a), in which a disorder potential with a characteristic lengthscale $\xi$ is added to the Ising Hamiltonian, and the resulting domain density fluctuations are blurred to reveal their superstructure, as for the experimental analysis (see Supporting Information Section S3 for details). As can be seen in \fref{Simulation}(b), we find that the typical feature size extracted from radially averaged autocorrelation analysis exactly matches the characteristic lengthscale $\xi$ of the disorder potential.

While the effect of strong, localised defects can be seen in, for example, the changes in roughness at domain walls \cite{Guyonnet--2013}, our observation of a characteristic lengthscale of 800 nm seems more compatible with variations of a collective background defect potential. This potential has been previously mapped out using spectroscopic techniques \cite{Jesse-NM-2008}, recently further assisted by machine learning \cite{Agar-AM-2018}. We note, moreover, that the key contributions to the disorder landscape do not appear to be variations of the surface morphology (see Supporting Information Fig.~S10 for details). Our analysis therefore demonstrates that quantitative information on this defect potential -- its characteristic lengthscale --  can be obtained using non-invasive and non-destructive optical measurements, without the need to switch the intrinsic domains and thus modify the defect landscape of the film \cite{Kalinin-AN-2011,Domingo-N-2019}. This method can be applied more generally, for example to samples where the defect density has been engineered, to reveal the resulting change in domain density superstructure.


In summary, our complementary PFM, SHG, and TEM measurements demonstrate that both artificial and naturally occurring domain walls in PbTiO$_3$ thin films present components of Néel polarisation at room temperature, possibly stabilised over the predicted Ising structure by defect segregation. Indeed, radially averaged autocorrelation analysis of domain density fluctuations reveals characteristic lengthscales which can be related to variations in the underlying disorder potential landscape. While it remains to be explored if and how such domain walls transition to a Bloch configuration at low temperature, our observations suggest potential utility in novel nanoelectronic applications, as Néel domain walls would be locally more likely to show increased electrical conductivity than their Ising counterparts (see Supporting Information Section 4 for details).

\section*{Methods} 
\paragraph{Thin film growth}
Three different samples have been used in this study:
\begin{itemize}
    \item M: monodomain sample studied by SHG
    \item P1: polydomain sample studied by TEM
    \item P2: polydomain sample studied by SHG
\end{itemize}

The samples were grown epitaxially using off-axis radio frequency magnetron sputtering on TiO$_2$ terminated, (001)-oriented, undoped SrTiO$_3$ with a LaNiO$_3$ back electrode. LaNiO$_3$ was deposited at a temperature of \SI{510}{\celsius} for samples M and P1 and \SI{555}{\celsius} for sample P2, in 180 mTorr of a 10:35 O$_2$/Ar mixture using a power of 50 W. The PbTiO$_3$ thin films were deposited at temperatures ranging from \SIrange{540}{555}{\celsius} (\SI{540}{\celsius}, \SI{545}{\celsius} and \SI{555}{\celsius} for sample M, P1 and P2 respectively) in 180 mTorr of a 20:29 O$_2$/Ar mixture using a power of 60 W. Additionally, the PbTiO$_3$ film in sample P2 was grown using the SKIS technique (Slow Kinetics Intermittent Sputtering, see~\cite{Weymann-ASS-2020})), while standard (continuous) RF growth was used for all the others. A Pb$_{1.1}$TiO$_3$ target with a 10 \% excess of Pb was used to compensate for Pb volatility, while stoichiometric targets were used for all other materials.

The different growth conditions have been used to optimize the sample quality and achieve the desired domain configuration (monodomain or polydomain) needed for this study. Note that this explains the slightly different vertical PFM phase images obtained on sample P1 (figure~\ref{Schematic}(a): mix of up and down regions) and on sample P2 (figure~\ref{SHGpoly}(a): uniform background with well defined nanoregions of the opposite polarization).

\paragraph{SPM Characterisation}
All SPM characterisation was performed using an {\it Asylum Research Cypher} atomic force microscope. Surface topographies were obtained in tapping mode using {\it Bruker TESPA} tips. Piezoresponse force microscopy vertical phase and amplitude were recorded in dual resonance tracking (DART) mode \cite{Dart2007} with excitation amplitudes typically less than 500 mV, using conductive {\it Mikromasch HQ:NSC18}/Cr-Au coated tips. 

\paragraph{SHG}
Local SHG measurements were conducted by means of an inverted optical microscope. The fundamental wave is provided by a laser source emitting pulses of 100 fs duration at a repetition rate of $80$ MHz, centered at a wavelength $\lambda = 800$ nm. The sample was illuminated at normal incidence with a time-averaged power of $10-14$ mW. The SHG images are obtained by scanning the sample with respect to the focused laser beam (objective x$40$, $0.66$ numerical aperture) using computer-controlled stepping motors. The output intensity was spectrally filtered and collected into a photomultiplier. Polarimetry measurements are performed by recording the images at different polarizer and analyzer angles (P and A, respectively). Ref. \cite{Cherifi-Hertel-NC-2017,Cherifi-Hertel-JAP-2021} provide detailed information on the measurement geometry, and on how the internal domain wall structure is determined using SHG polarimetry.

\paragraph{TEM}
Cross-sections for STEM were prepared via FIB milling at accelerating voltages of 30, 16, 8, 5 and 2 kV on an FEI {\it Scios DualBeam} instrument. STEM measurements were performed on a probe-corrected  FEI {\it Titan Themis} operated at 200 kV and with a probe convergence angle of 21.2 mrad. High angle annular dark field (HAADF) imaging was performed with inner/outer collection angles of 56.3 and 200 mrad, respectively. DPC imaging was performed with a 4 segment annular detector with inner/outer collection angles of 13.2 and 73.6 mrad, respectively. The specimen thickness was estimated from the $t/\lambda$ (thickness/inelastic mean free path) from a zero loss electron energy loss (EEL) spectrum acquired using a {\it Gatan Enfinium} spectrometer. Atomic resolution images were acquired as series of 15 frames with pixel dwell time of 100 ns before being summed and corrected for drift in Velox by {\it Thermo Fisher Scientific}. Gaussian fitting was performed using Atomap \cite{Nord-ASCI-2017} and Matlab.

\section*{Data availability}
The data that support the findings of this study are available at Yareta (DOI).

\section*{Acknowledgements}
This work was supported by Division II of the Swiss National Science Foundation under project 200021\_178782. A.B.N. greatfully acknowledges support from the EPSRC (EP/R023751/1 and EP/L017008/1). S.C.-H. acknowledges the support of the French Agence Nationale pour la Recherche (ANR) through grant ANR-18-CE92-0052 and the Programme d’Investissement d'Avenir ANR-10-IDEX-0002-02 of the University of Strasbourg under contract ANR-11-LABX-0058\_NIE. C.L. acknowledges the support from the Division II of the Swiss National Science Foundation under project 200021\_200636. The authors thank S. Mohapatra for his help for the domain poling. 

\section*{Author contributions}
P.P. and S.C.-H. designed the experiment. C.W. and C.L. grew the samples and conducted the piezoresponse force microscopy. S.C.-H. carried out the second harmonic generation microscopy in collaboration with K.D.D. A.B.N. performed the transmission electron microscopy and the atomic position fitting. C.W. carried out the auto-correlation analysis. I.G. performed the Monte Carlo simulations of the perturbed Ising Model. C.W., C.L., A.B.N., and P.P. wrote the manuscript with contributions from all authors. All authors discussed the experimental results and models, commented on the manuscript, and agreed on its final version.


\end{document}


\maketitle

\renewcommand{\thepage}{S\arabic{page}} 
\renewcommand{\thesection}{S\arabic{section}}  
\renewcommand{\thetable}{S\arabic{table}}  
\renewcommand{\thefigure}{S\arabic{figure}}

\section{STEM Data Analysis}

The atomic resolution images in the manuscript are the result of drift corrected frame imaging (DCFI) whereby multiple frames with low pixel dwell time are added together and drift corrected to avoid issues of sample/stage drift. In this work, we used 15 frames of 2048$^2$ pix$^2$ with dwell time of 200 ns/pixel.

In order to extract the average positions of the Pb and Ti atom columns, 2D Gaussians were fitted using Atomap following the procedures detailed elsewhere \cite{Nord-ASCI-2017}. \fref{InitialSTEMfitSupp}(a) shows the drift-corrected HAADF image discussed in the main paper Fig.~3 with the fitted Pb and Ti atom positions overlaid in red and green, respectively. The corresponding DPC image, acquired simultaneously, is shown in \fref{InitialSTEMfitSupp}(b) where the ``kink'' discussed in the main paper (Fig.~3) can be seen as the vertical yellow line in the centre, arising from the shifts in atom positions there.

\subsection{Ti and Pb ellipticity}\label{sec:Ellipticity}

From the drift-corrected HAADF image in \fref{InitialSTEMfitSupp}(a), one can also determine the ellipticity of the atom columns, which gives a measure of the disorder through the thickness of the lamella. Indeed, an ideally ordered column of atoms would appear perfectly circular. The ellipticity of the Ti atom columns is highly directional along the c-axis, resulting in the intensity being localised to a central point with streaking occurring either side. 

One has to be careful as scan/sample/stage drift would result in either a uniform smearing out of the atom column intensity (identical for both Ti and Pb) if the drift were highly directional, or there would be an apparent enlargement of the atom columns were the drift random and non-directional. We didn't have such drift in our measurements.

The Pb atoms are also likely to be influenced by the STEM probe due to the high vapour pressure (volatility) of Pb. We did not observe any variation of atom column intensities across the 15 acquired frames, excluding the possibility of knock-on damage causing sputtering of Pb or Ti. We can therefore rule out the various forms of drift and damage as being the cause of the observed Ti ellipticity.

\subsection{Initial fitting based on Pb and Ti positions}\label{sec:InitialFitting}

\begin{figure}[htbp]
	\centering
	\includegraphics[width=0.8\textwidth]{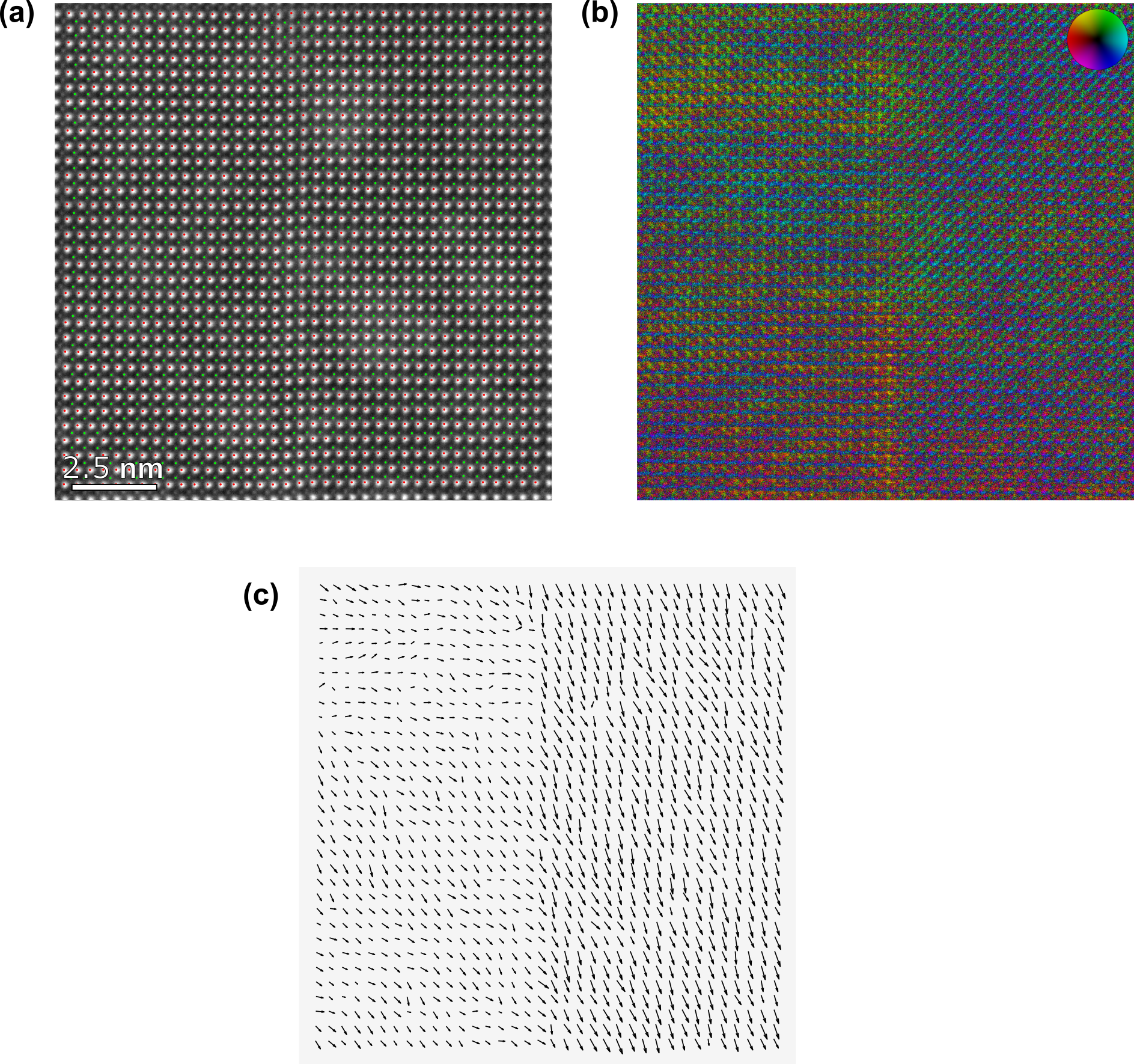}
	\caption{\emph{Initial fitting based on Pb and Ti positions.} (a) Atomic resolution HAADF image shown in the main text with Pb and Ti fitted atom column positions overlaid in red and green, respectively. (b) Corresponding DPC image and (c) polarisation determined from the Ti displacement from the centre of the unit cell defined by the nearest Pb atom columns. The scale bar is the same for all images.}
	\label{InitialSTEMfitSupp}
\end{figure}

Using the Ti displacement from the centrosymmetric position defined by the nearest Pb atom columns, we can extract an estimate of the local polarisation within each unit cell, shown in \fref{InitialSTEMfitSupp}(c). As shown in ab-initio calculations by Bousquet {\it et al}~\cite{Bousquet2008}, for uniaxial ferroelectric distortions along the c-axis orientation, with PbTiO$_3$ in the \textit{P4mm} phase, and for relaxation of the atomistic simulations performed at constant volume, the Ti displacement vector is oriented in opposite direction to the polarisation and related to the latter via $P_\alpha = \frac{e}{\Omega} \Sigma_{j\beta}Z^*_{j\alpha\beta}\Delta u_{j\beta}$ where $\Delta u_{j\beta}$ is the displacement of ion $j$ in the direction $\beta$, $Z^*_{j\alpha\beta}$ is the Born effective charge tensor, and $\Omega$ is the unit cell volume~\cite{Neaton-PRB-2005,Ghosez-PRB-1998}. The resulting polarisation estimate shows a largely uniform downwards orientation on the right hand side of \fref{InitialSTEMfitSupp}(c), whereas at the ``kink'' in the centre and on the left hand side there is a clearly reduced out-of-plane polarisation component and increased in-plane component.

\subsection{Adding the oxygen position}\label{sec:InitialFittingOxygen}

For a more accurate estimate of the polarisation in PbTiO$_3$, the position of the oxygen octahedra should also ideally be taken into account.

\begin{figure}[htbp]
	\centering
	\includegraphics[width=0.8\textwidth]{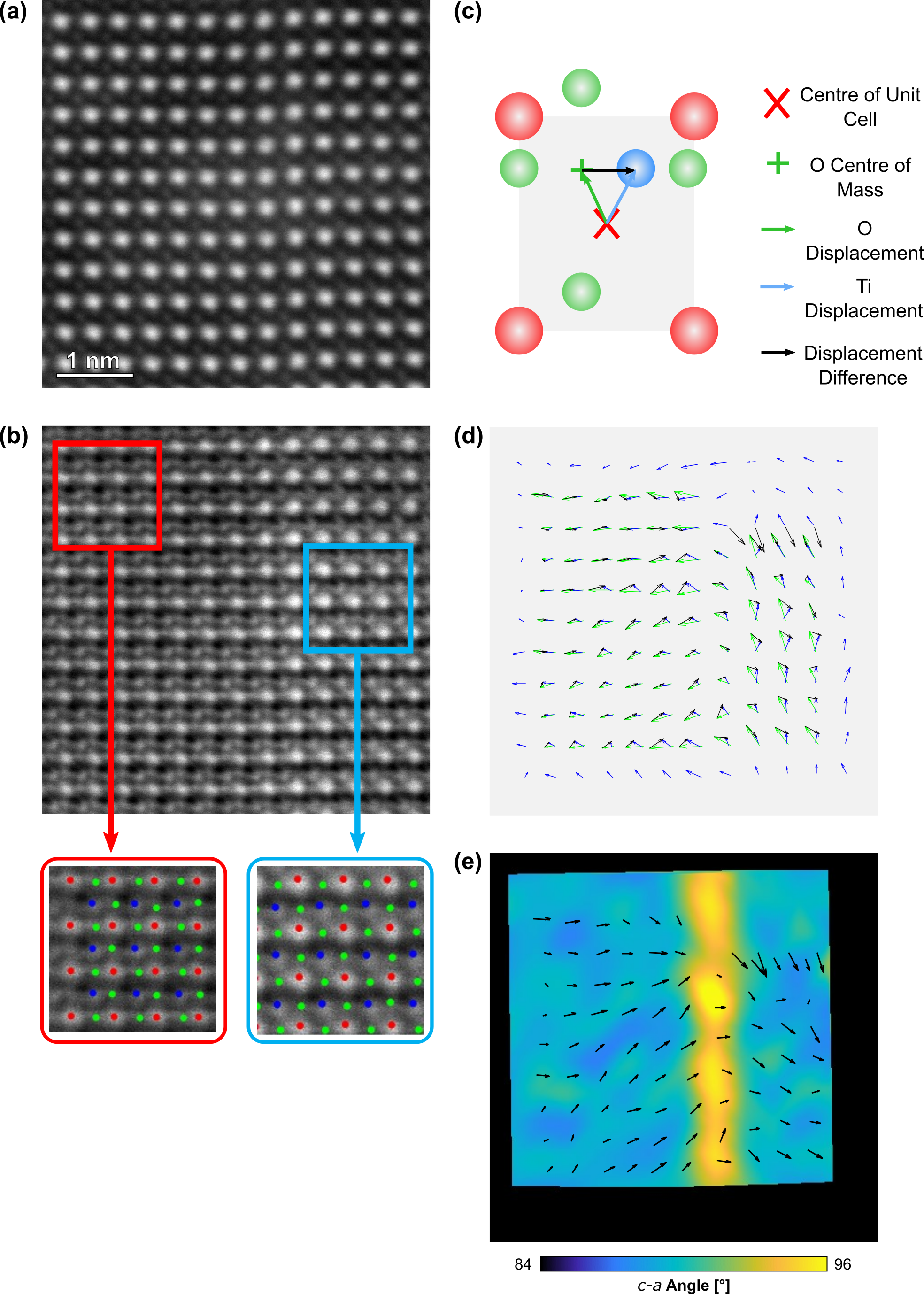}
	\caption{\emph{Oxygen displacement} (a) Higher magnification HAADF image across the same region as in \fref{InitialSTEMfitSupp}, (b) corresponding ABF image (shown as the reciprocal to ease identification of atom columns), the insets show enlarged regions of the image with the fitted atom positions overlaid (Pb in red, Ti in blue and O in green). (c) Schematic depicting definition of the different atomic displacements, (d) O (green), Ti (blue) and displacement difference (black) as measured from (a,b) and (e) displacement difference overlaid on the $c-a$ angle. The scale bar is the same for all images.}
	\label{STEM-OdisplacementsSupp}
\end{figure}

In order to better approximate the polarisation, Fig~\ref{STEM-OdisplacementsSupp}(a,b) shows higher magnification HAADF and annular bright field (ABF) images, respectively, of the same region as \fref{InitialSTEMfitSupp}. From the insets of \fref{STEM-OdisplacementsSupp}(b), it can be seen that there are subtle differences in the O positions (green dots) relative to the unit cell defined by Pb (red dots) when looking either side of the ``kink''. To map these variations, we define the O displacement as the shift of the centre of mass of the 4 nearest neighbour O atoms from the centre of the unit cell, in the same manner as for the Ti displacement discussed in the main text, and as schematically depicted in \fref{STEM-OdisplacementsSupp}(c). We can then extract the difference between the O centre of mass and the nearest Ti atom position, or ``displacement difference'', which represents a more accurate approximation of the polarisation than the Ti displacement alone. The results of this approach are shown in \fref{STEM-OdisplacementsSupp}(d) with O and Ti displacements in green and blue, respectively, and the displacement difference in black. \fref{STEM-OdisplacementsSupp}(e) shows the O-Ti displacement difference (i.e. polarisation) overlaid on the $c-a$ angle determined from \fref{STEM-OdisplacementsSupp}(a). These results illustrate that, although there is a noticeable difference between the O-Ti displacement and the Ti displacement, both metrics display clear Néel-like behaviour across the domain wall.

\subsection{Advanced fitting based on 3D Gaussian}

These observations, coupled with the variations in tetragonality, $c-a$ angle and ellipticities of the atom columns shown in the main paper Fig.~3, suggest the presence of a domain wall. We note that in particular at the domain wall, the unit cells are clearly distorted, and the Born effective charge is therefore likely different from that of a bulk domain region. Moreover, the real polarisation is blurred out due to projection effects, as suggested in Fig.~2. 

In order to extract the different contributions to this projected polarisation, we fit three 2D Gaussians to each of the Ti atom columns, using the Gaussian parameters from the initial fitting as the starting point. Three contributions are chosen to reflect opposing Ti atom shifts up and down relative to this mean position and the third to account for the ``mean'' position between the up/down ones. However, it is first necessary to carefully define a set of restrictions to limit the degrees of freedom such that the results are physically meaningful since almost any function can provide a reasonable fit when provided with sufficient terms.

Firstly, we only apply this three Gaussian fit if the Ti ellipticity exceeds a value of 1.1 since a small value indicates a well-ordered column of atoms. The value of 1.1 is chosen based on fits to the SrTiO$_3$ substrate. The ellipticity $\epsilon$ is defined as:
\begin{equation}
  \epsilon =
    \begin{cases}
      \sigma_x / \sigma_y, & \text{if $\sigma_x$ > $\sigma_y$}\\
      \sigma_y / \sigma_x, & \text{if $\sigma_y$ > $\sigma_x$}
    \end{cases}       
\end{equation}
where $\sigma$ are the Gaussian standard deviations along the $x, y$ directions given by the subscripts. Secondly, these values of standard deviation should not be allowed to become excessively large or small since this would reflect an unrealistic fit due to, for example, background intensity variations. A reasonable limit can be determined in this system by using the standard deviations obtained from the SrTiO$_3$ single crystal substrate as a baseline. Here we use the $\sigma$ limits for Ti of $\sigma_{PTO} = \sigma_{STO} \pm 5$ pixels, corresponding to $\sim$ 36 pm here (remembering that these $\sigma$ values are the results of convolution of the atom columns with the STEM probe). Finally, the three Ti positions should be restricted such that excessively large displacements do not occur, i.e. so the extracted atom positions do not exit the unit cell, which would be unphysical. In other words, if the Ti position from the initial fit is close to the top/bottom of the unit cell, then we restrict the up/down Gaussian displacement in the multiple fitting. Here, we employ a limit of $\sim \pm$ 40 pm based on our previous studies of monodomain PbTiO$_3$ films \cite{Weymann-AEM-2020}.

\begin{figure}[htbp]
	\centering
	\includegraphics[width=0.8\textwidth]{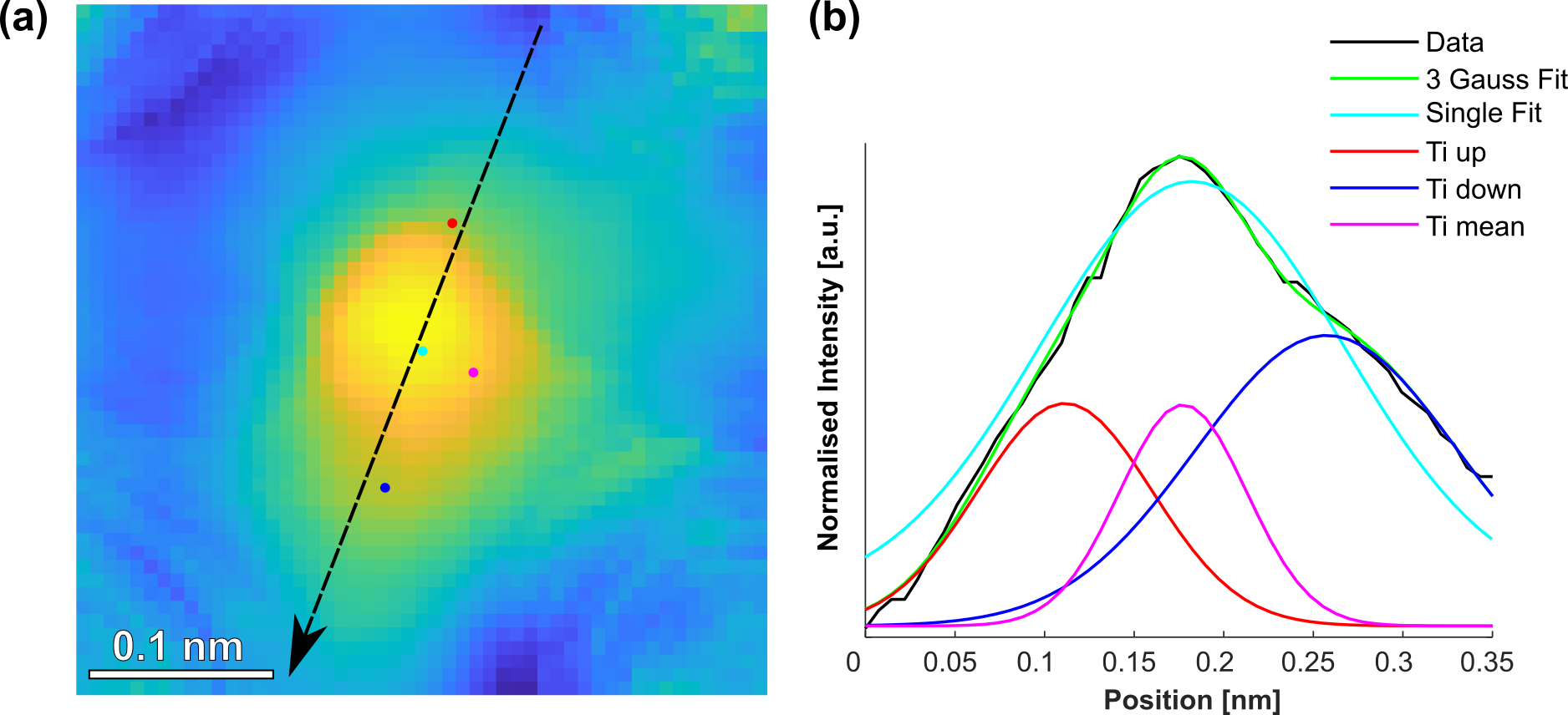}
	\caption{\emph{3 Gaussian fitting of a single Ti atom column.} (a) A cropped portion of the HAADF image from \fref{STEM-OdisplacementsSupp}(a) after subtraction of the Pb atoms, showing the remaining Ti atom column in the centre. The refined positions of the 3 Gaussians fitted to the Ti atom column are overlaid (red, magenta and blue circles) along with the original 1 Gaussian fit shown in cyan in the centre. (b) Line profile of (a) taken along the black dashed line, showing the contributions of the different components with the colours matching those in (a). `Data' refers to the intensity data extracted from (a) and `Ti up/down/mean' refer to the components from the 3 Gaussian fitting.}
	\label{3GaussDemoSTEMfitSupp}
\end{figure}

\fref{3GaussDemoSTEMfitSupp} shows an example of the result of 3 Gaussian fitting to a single Ti atom column. In \fref{3GaussDemoSTEMfitSupp}(a) we present the HAADF image after subtraction of the Pb atom columns so that only the B-site Ti atoms remain; it is plotted with a colour map rather than greyscale to draw attention to the more subtle intensity variations. The overlaid circles show the locations of the refined atom positions: cyan is the original 1 Gaussian fit and shown in red, blue and magenta are the positions from the 3 Gaussian fit. It can be difficult to visualise the subtleties from a 2D image, therefore \fref{3GaussDemoSTEMfitSupp}(b) shows line profiles taken along the direction of ellipticity, shown by the black dashed arrow in \fref{3GaussDemoSTEMfitSupp}(a). From these profiles, it can be clearly seen that while the single fit (cyan) accurately captures the centre of mass of the atom column (black), the former does not accurately model the more subtle intensity variations that cannot be faithfully reproduced with a single Gaussian. On the other hand, the 3 Gaussian fit (green), shows a much closer match to the original. This fit is the sum of the three Gaussians shown in magenta, red and blue corresponding to a central, or ``mean'' Ti position and relative shifts up and down (red and blue, respectively). In other words, the Ti atom columns observed can be approximated to result from the superposition of three differing polarisation states.

\begin{figure}[htbp]
	\centering
	\includegraphics[width=0.8\textwidth]{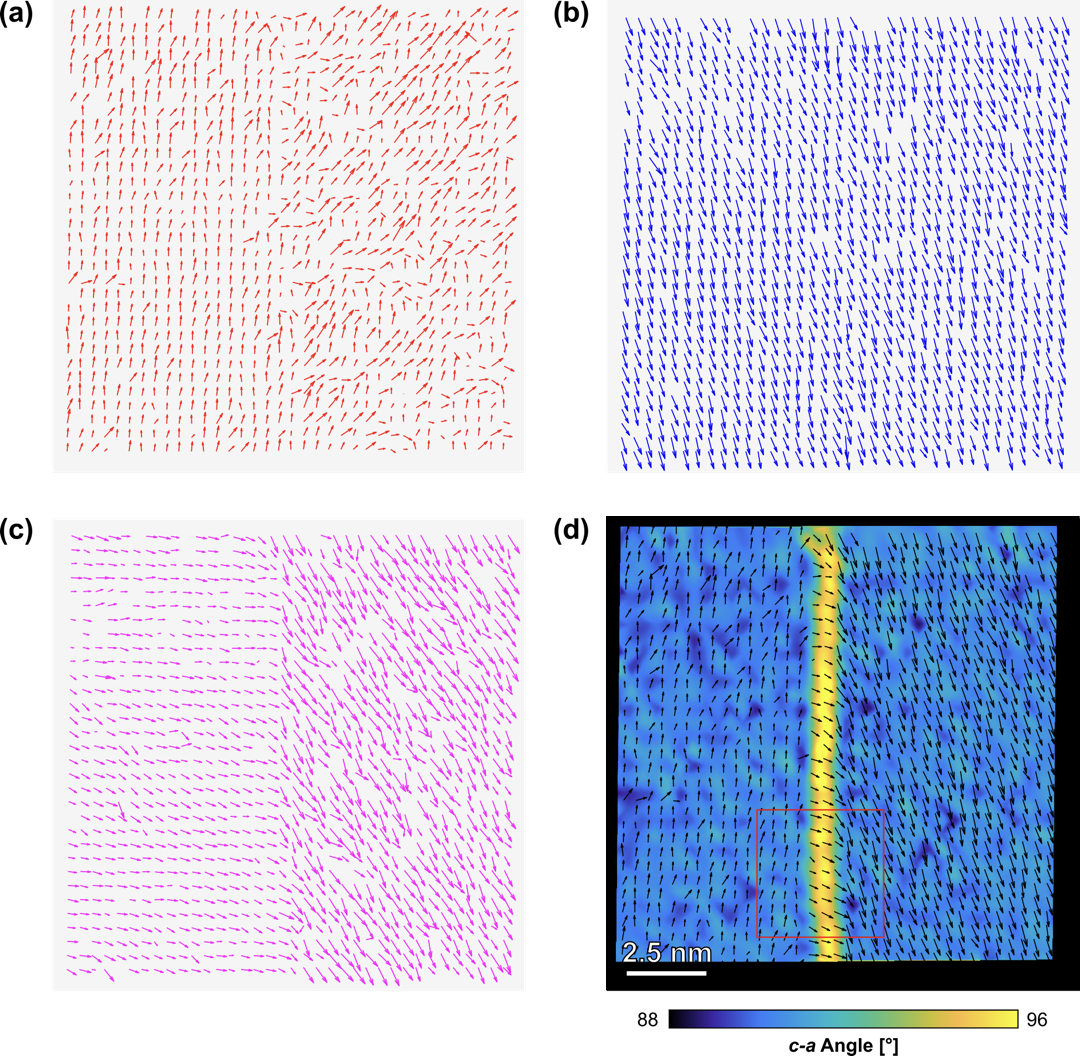}
	\caption{\emph{Polarisation components and reconstruction without projection.} Polarisation determined from the Ti positions given by the three Gaussian fitting corresponding to (a) Ti up, (b) Ti down and (c) Ti mean. The colours match those shown in \fref{3GaussDemoSTEMfitSupp}. (d) Polarisation after reconstruction to remove projection effects overlaid on the $c-a$ angle.}
	\label{3GaussResultsSTEMfitSupp}
\end{figure}

We now turn to the full results from this fitting: \fref{3GaussResultsSTEMfitSupp}(a)--(c) shows the polarisation for the different components resulting from the 3 Gaussian fitting across the full image. \fref{3GaussResultsSTEMfitSupp}(a) shows a predominantly upward polarisation with very small in-plane components on the left hand side whereas much larger and more variable in-plane components can be seen on the right. In contrast, \fref{3GaussResultsSTEMfitSupp}(b) shows uniform downward polarisation with little variation across the image. From \fref{3GaussResultsSTEMfitSupp}(c), the mean Ti displacement gives a mainly in-plane polarisation on the left hand side, which rotates to have greater down-oriented out-of-plane components on the right hand side. Based on these observations and those from the main text and section \ref{sec:InitialFitting}, we can draw two main conclusions to aid reconstruction of the overall polarisation in the absence of projection effects:
\begin{enumerate}
    \item at the apparent domain wall, there is an increase in the $c-a$ angle accompanied by a decrease in the c-axis lattice parameter. This suggests greater in-plane polarisation across this apparent wall, as observed in \fref{3GaussResultsSTEMfitSupp}(c), and
    \item from the single Gaussian fitting, the out-of-plane Ti displacement is smaller on the left side of the apparent wall and larger on the right, suggesting greater influence of overlapping polarisation on the left.
\end{enumerate}

As such, it is reasonable to conclude that the upwards polarisation observed in \fref{3GaussResultsSTEMfitSupp}(a) is more important on the left, and the down polarisation in \fref{3GaussResultsSTEMfitSupp}(b) is more important on the right, with the strong in-plane polarisation from \fref{3GaussResultsSTEMfitSupp}(c) occurring at the apparent domain wall in the centre. Thus, in order to reconstruct the polarisation, we iterate through each atom column to set the polarisation as resulting from `Ti mean' in \fref{3GaussResultsSTEMfitSupp}(c) where the $c-a$ angle is $>93^{\circ}$ then being due to either `Ti up' (\fref{3GaussResultsSTEMfitSupp}(a)) to the left and `Ti down' (\fref{3GaussResultsSTEMfitSupp}(b)) on the right. \fref{3GaussResultsSTEMfitSupp}(d) shows the result of this reconstructed polarisation overlaid on the interpolated $c-a$ map. In other words, the left side of \fref{3GaussResultsSTEMfitSupp}(d) results from \fref{3GaussResultsSTEMfitSupp}(a), the centre -- where the angle is large -- from \fref{3GaussResultsSTEMfitSupp}(c) and the right hand side from \fref{3GaussResultsSTEMfitSupp}(b). The remaining or ``residual'' polarisation components from \fref{3GaussResultsSTEMfitSupp}(a)-(c) therefore constitute the variations of polarisation through the thickness of the cross-section.

\section{Radially averaged autocorrelation analysis} 

The radially averaged autocorrelation measures the total correlation between an image and its copy shifted by a certain distance in any direction. This allows us to extract both the characteristic feature size for each image, which is given by the position of the first minimum in the radially averaged autocorrelation curve, and the pseudo-period of any superstructure in the image, which is given by the position of the first maximum in the curve.

\subsection{Extracting the radially averaged autocorrelation in PFM and SHG images}

To quantify the contrast variations in the PFM and SHG observations, we ana\-lysed their radially averaged autocorrelation to extract the characteristic feature size and pseudo-period. As can be seen in the main paper Fig.~4(e), while the SHG image shows no clear characteristic feature size, possibly as a result of the high heterogeneity of the features, it does appear to have a pseudo-period of approximately \SI{800}{nm}. The origin of these SHG signal variations can be explained by considering the cumulative optical response of many domain walls in close proximity. In an area where the domain density is uniformly high or low, opposing contributions from neighbouring domain walls will cancel out. However, in the transition region from high to low domain density, this compensation will no longer be perfect, and we should detect some SHG signal.

From the PFM phase image of the domains, on the other hand, we extract a pseudo-period of about \SI{100}{\nano\metre}, which corresponds to the domain structure, and no further marked peaks which could indicate a periodicity in the domain density variations. The results for the domain walls, extracted from the phase image using a Canny edge detection algorithm, are even more extreme: beyond the scale of a domain, the correlation is consistently very close to $0$.

To access the variations in the local density of domains and domain walls, and compare the PFM measurements with the observations made by SHG, we explicitly blurred out the microstructure in the PFM images by convoluting the phase and amplitude images with a Gaussian kernel with a standard deviation of $\sigma=\SI{100}{\nano\meter}$, larger than any individual domain, and a size of at least $3\sigma$ in all directions to ensure a smooth result. This is equivalent to replacing each point in the image with a Gaussian weighted average of all of the pixels in its neighbourhood. This blurring process clearly reveals variations in domain density, both when applied directly to the PFM phase imaging (main paper Fig.~4(c)) and to the domain wall mapping (main paper Fig.~4(d)), with the resulting bright/dark contrast variations appearing remarkably similar to the superstructure visible in SHG (main paper Fig.~4(b)). More quantitatively, the radially averaged autocorrelation of this resulting superstructure shows the same pseudo-period as that of the intensity variations in the SHG image Fig.~4(e). 

\subsection{Validation of the radially averaged autocorrelation approach for PFM and a discussion of possible artefacts}

To validate the statistical analysis of the domain superstructure, we verified that both domain size and periodicity can be reliably extracted from simulated domain structures using the radially averaged autocorrelation, and that the choice of blurring radius does not affect the extracted periodicity. Furthermore, we verified that common sources of artefacts in PFM images, namely the resolution and the orientation of the fast scan direction, do not affect our results. These different potential artefacts are discussed in detail below.
	
\begin{figure}[hp]
	\centering
	\includegraphics[width=\textwidth]{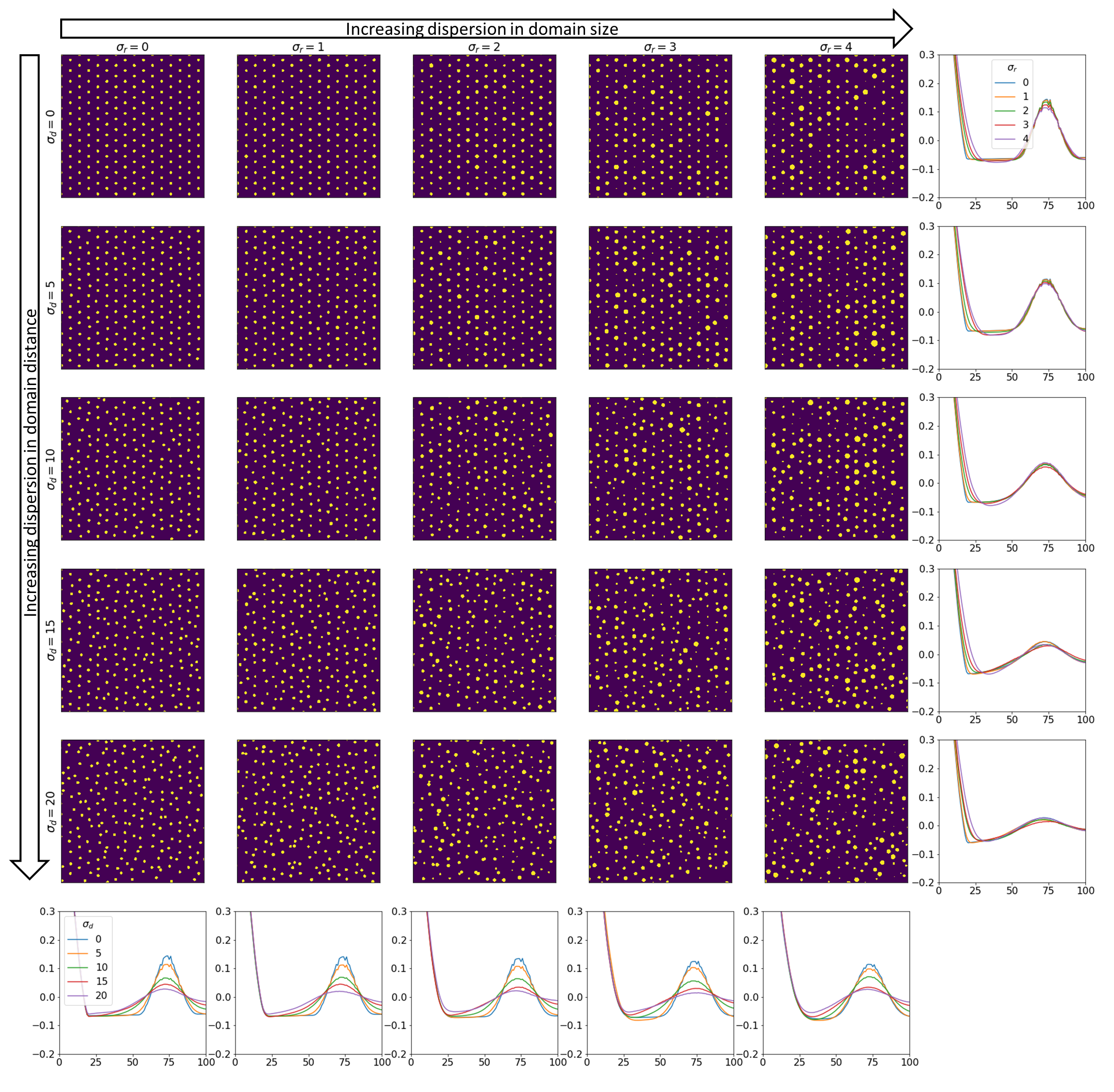}
	\caption{\emph{Radially averaged autocorrelation for a regular grid of circles.} The average radius of one circle is $\mu_r=10$ and the average separation between neighbouring circles is $\mu_d=75$. Each circle was displaced in a random direction by a distance drawn from a normal distribution with standard deviation $\sigma_d = 0,5,10,15,20$ along each line respectively, while each radius was drawn from a normal distribution with standard deviation $\sigma_r=0,1,2,3,4$ along each column respectively. The graph at the end of each line/column shows the radially averaged autocorrelation for each configuration with this particular disorder in distance/radius. We can see that the typical feature size (the circle diameter), given by the position of the first minimum in autocorrelation, is hard to determine as soon as disorder is introduced. However, the position of the first maximum in the autocorrelation, giving the typical distance between two features, is extremely robust under both types of disorder.}
	\label{Disorder}
\end{figure}

\fref{Disorder} shows simulated domain patterns with various degrees of disorder, introduced both as variations of the radius of each circular domain ($\sigma_r$) and as variations in the distance between domains ($\sigma_d$), while keeping the average of these values constant. Focusing on the effect of disorder in the domain size, which increases from left to right and is summarised in the radially averaged autocorrelation graphs in the rightmost column, we can see that, when compared to a perfect pattern ($\sigma_r=0$, left column), the position of the first minimum shifts from exactly $20$, the diameter of one of the domains, towards the midpoint between $0$ and the first maximum. Indeed, an increasingly wider distribution of domain sizes extends the tail of both the central peak around $0$ and the peak around the first local maximum. Since it is this peak which indicates the period, this type of disorder makes it hard to accurately extract the typical domain size if it extends to values beyond half the periodicity.

Comparable disorder perturbation of the periodicity, however, shows that the autocorrelation technique is much more robust in this regard. Along the columns of \fref{Disorder}, we show domain patterns where each domain has been moved from its position in the perfect triangular lattice in a random direction, by an amount drawn from a normal distribution with increasing standard deviation up to $20$, corresponding to \SI{\approx25}{\percent} of the distance between the perfect ``lattice sites''. Again, the bottom line of \fref{Disorder} shows the resulting radially averaged autocorrelation. Although the first peak diminishes in intensity and becomes wider with increasing disorder, its position reliably indicates the average period of the domain structure, $75$.

\begin{figure}[htbp]
	\centering
	\includegraphics[width=0.8\textwidth]{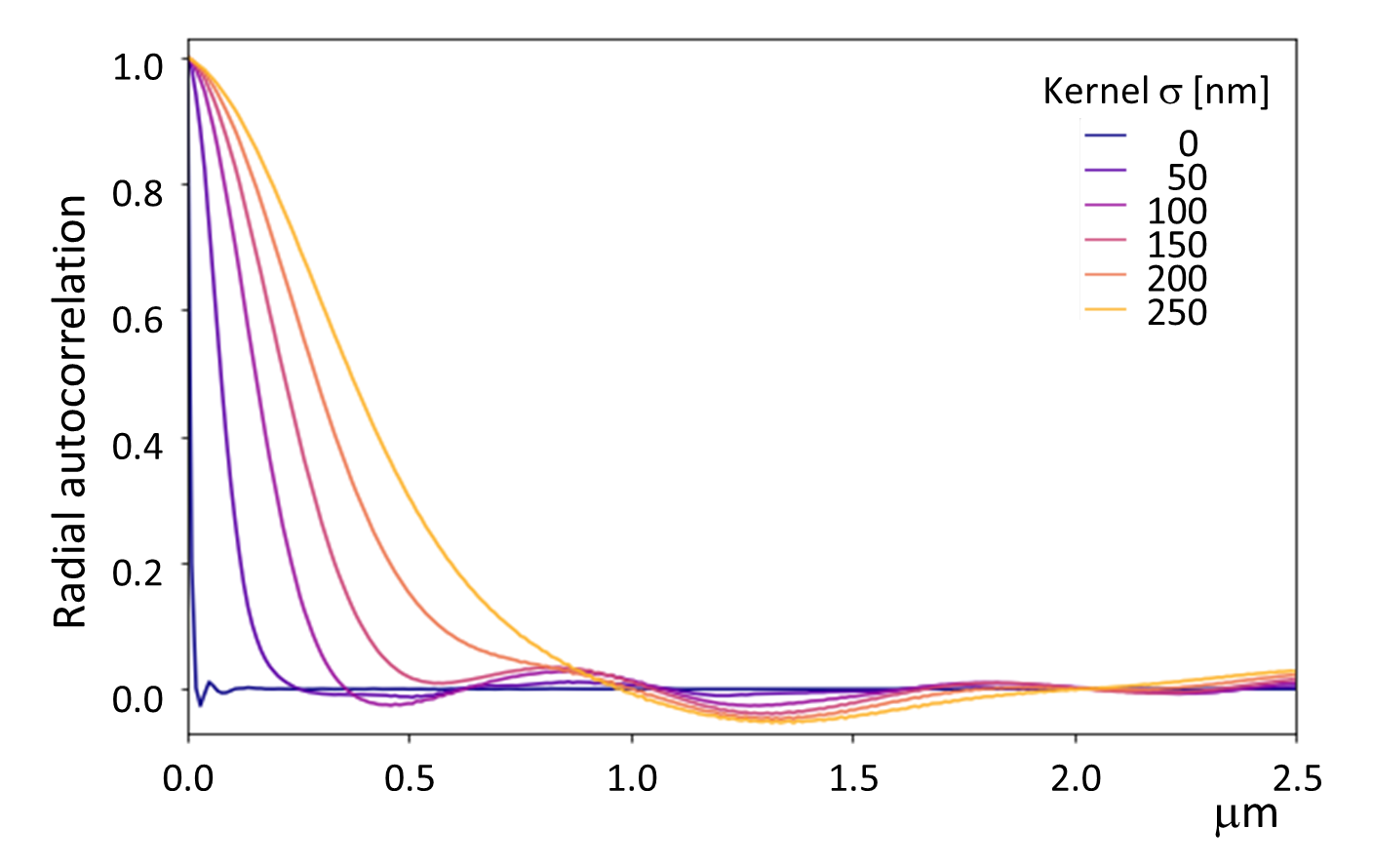}
	\caption{\emph{Effect of blurring.} Radially averaged autocorrelation of the density of edges (domain walls) in the PFM image shown in the main paper Fig.~4(a), (d), blurred by using Gaussian kernels of varying standard deviation to obtain the density superstructure. We can see that the pseudo-period of the density variations extracted using this method does not depend on the standard deviation of the kernel.}
	\label{BlurringRadius}
\end{figure}

Another important check is if the blurring of the microstructure introduces any spurious lengthscale into the autocorrelation function. To test this, we blurred the domain edge structure shown in Fig.~4(a), (d) with Gaussian kernels of various sizes before taking the radially averaged autocorrelation. The results are shown in \fref{BlurringRadius}: above the threshold of the domain size, the same superstructure period is extracted independently of the size of the kernel. The kernel is however convoluted with the central peak and the first peak, making them wider for larger kernels, and making the determination of a typical feature size in the superstructure challenging. As such, a sensible upper limit to the size of the Gaussian kernel is necessary. Nonetheless, our statistical tool is adequate to extract a typical period in the density variations of the domain structure of our film.

\begin{figure}[htbp]
	\centering
	\includegraphics[width=0.8\textwidth]{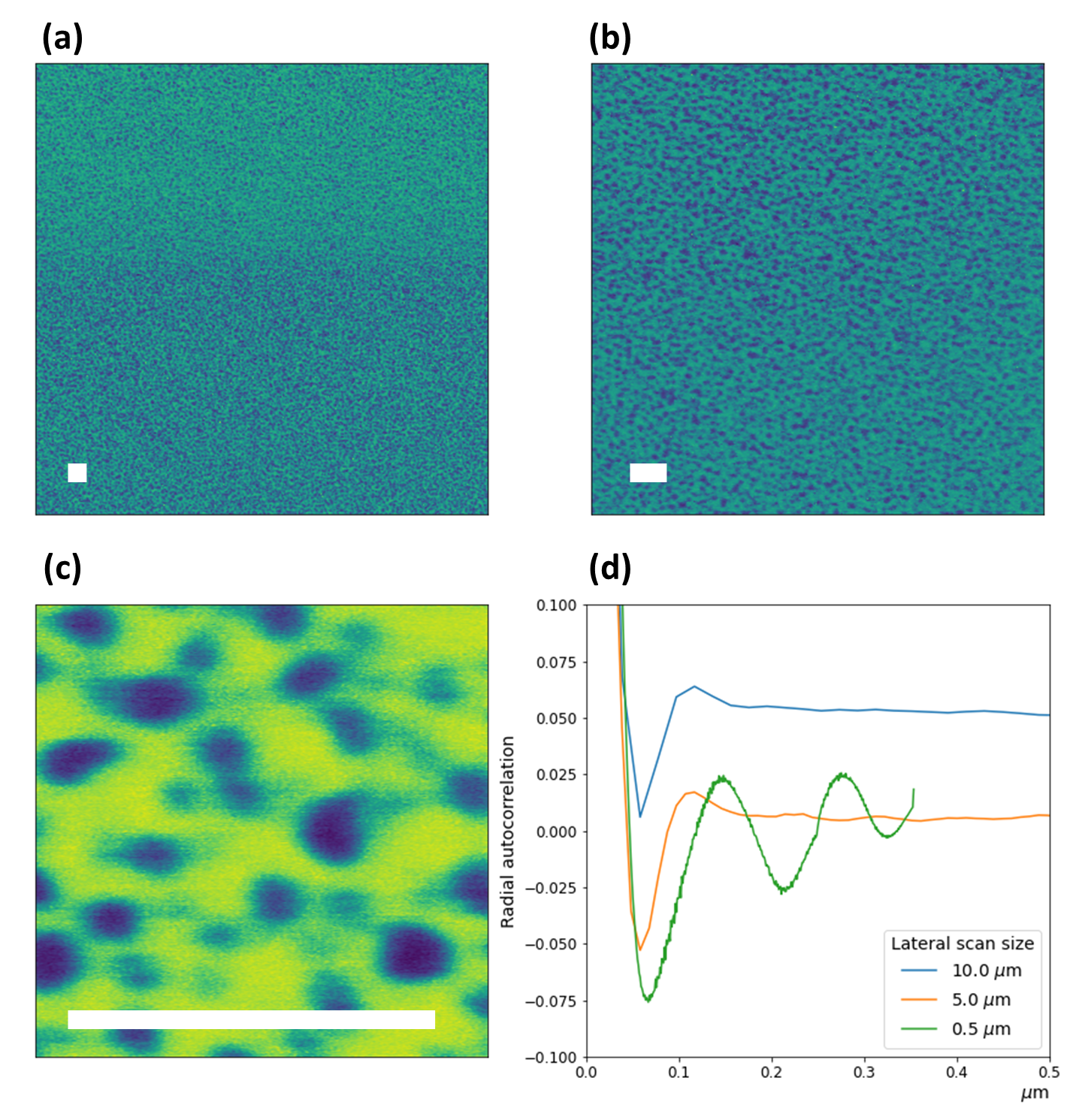}
	\caption{\emph{Effect of PFM resolution.} PFM phase image of the same region of one of our samples imaged with a lateral scan size of (a) \SI{10}{\micro\meter} (b) \SI{5}{\micro\meter}, and (c) \SI{0.5}{\micro\meter}, along with (d) the radially averaged autocorrelation for each of these images. The white scalebar in (a)--(c) represents \SI{400}{\nano\meter}. Each scan image contains 512 lines of 512 points. As can be seen in (d), this means that the $10 \times 10$~\si{\micro\meter}$^2$ image lacks the resolution to properly track the domain period, while the $500 \times 500$~\si{\nano\meter}$^2$ image seems to lack sufficient statistics to find the right average domain period.}
	\label{Resolution}
\end{figure}

Beyond the nature of the disorder and the effects of blurring protocol, size and resolution effects are known sources of artefacts during PFM imaging. In \fref{Resolution}, we analyse how they affect our statistical analysis technique. \fref{Resolution}(d) shows the radially averaged autocorrelation for each of the three PFM images of the domain structure of one of our samples (\fref{Resolution}(a)--(c)), imaged in the same region, but with varying lateral size and resolution. We can see that the curve obtained from the largest field of view image only has a few points in the first minimum and maximum, making it challenging to accurately extract the feature size and periodicity. The smallest field of view image, on the other hand, appears to contain too few domains to allow reliable extraction of the true average domain period, leading to a shifted first maximum. This effect would be further compounded by applying blurring to look at the superstructure, since the lengthscales of interest are about 5 times larger in that case. The results shown in the main paper therefore only use $5 \times 5$~\si{\micro\meter} images.

\begin{figure}[htbp]
	\centering
	\includegraphics[width=\textwidth]{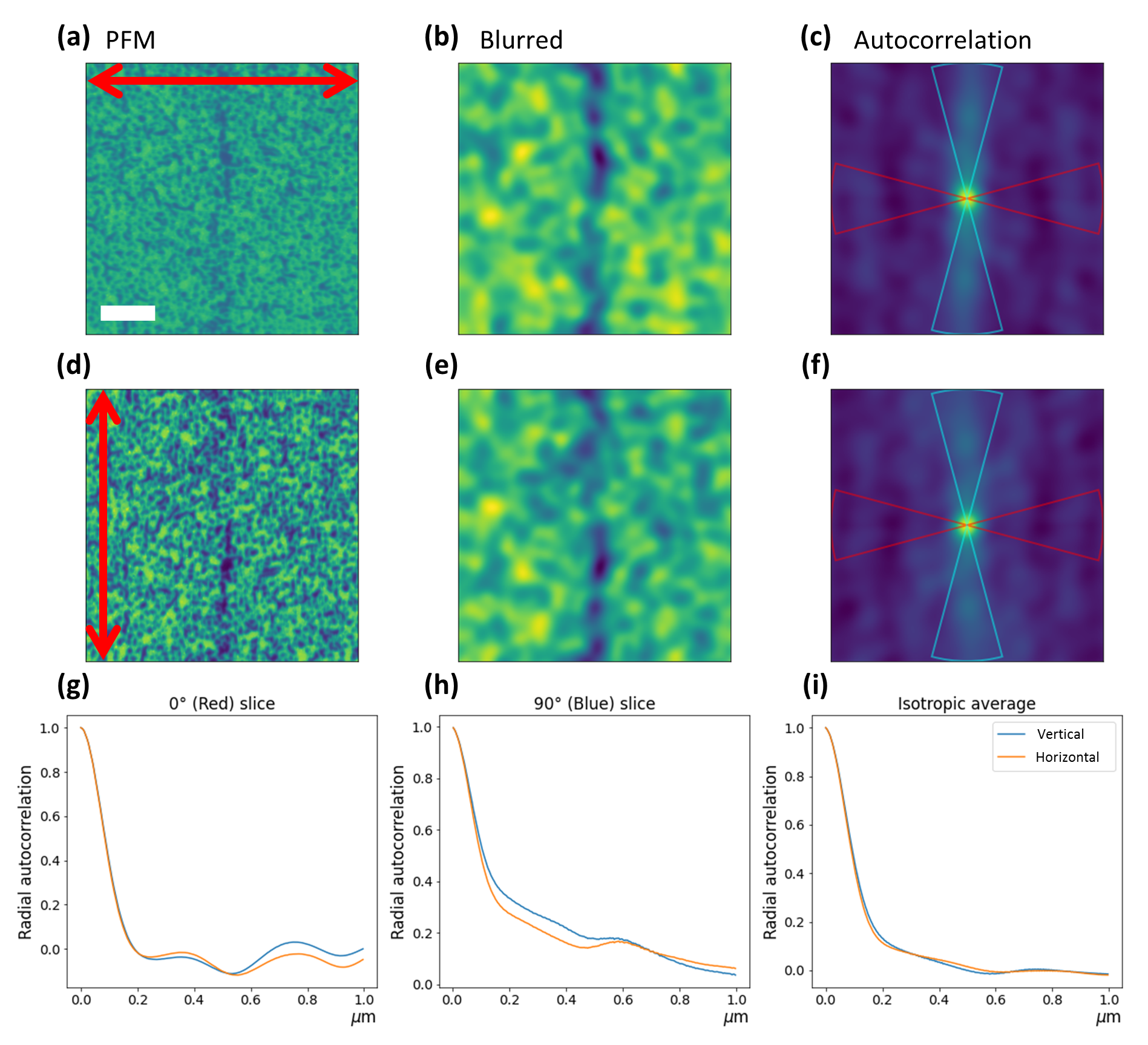}
	\caption{\emph{Effect of scanning direction.} Raw (a, d) and blurred (b, e) PFM phase images, and the corresponding autocorrelation (c, f) of the domain density superstructure extracted after blurring for two images of the same region scanned with the fast scan axis oriented horizontally (a)--(c) and vertically (d)--(f), as represented by the red arrow. The white scalebar is valid for all images and represents \SI{400}{\nano\meter}. The graphs in (g)--(i) show the autocorrelation radially averaged over the red (g) and blue (h) slices indicated in (c, f), and isotropically over all directions (i). The presence of a strongly directional defect in the domain structure in (a, d) is amplified by blurring (b, e), and significantly affects the period extracted along the vertical direction (h), leading to an unusable isotropic average (i). To avoid effects such as this, one must not only consider the isotropic radial average, but also the radial average in different directions.}
	\label{ScanDirection}
\end{figure}

Another common source of artefacts in scanning probe microscopy images is the asymmetry between the fast and slow scan directions, which typically manifests as `streaks' along the fast scan axis. In \fref{ScanDirection}, we consider two images of the same region oriented with orthogonal fast scan directions. Note that in this region, there is a real defect in the domain structure, which therefore appears in the same direction independently of the choice of fast scan axis orientation. This defect becomes even more prominent after blurring, and leads to some anisotropy in the autocorrelation. We can reveal this anisotropy by limiting our radial average to a section of the plane oriented in a specific direction. These sections are shown in red and blue on \fref{ScanDirection}(c, f), and the corresponding radial averages in \fref{ScanDirection}(g, h). The shift in the position of the minima in these two sub-averages leads to the fact that the full average, shown in \fref{ScanDirection}(i), does not exhibit a strong minimum or maximum, making it impossible to extract the feature size and the period. Here, the anisotropy is actually present in the underlying image, as evidenced by the fact that the positions of the minima and maxima of the different curves do not depend on the fast scan axis, but it could also be introduced by the choice of fast scan direction. It is therefore important to keep such anisotropies in mind while analysing scanning probe microscopy images using this technique. None of our other images shown in the main paper or the supplementary material presented such anisotropy, which is why we are always able to safely use the full (isotropic) radial average of the autocorrelation.

\section{Ising model with disorder}
\begin{figure}[htbp]
	\centering
	\includegraphics[width=0.8\textwidth]{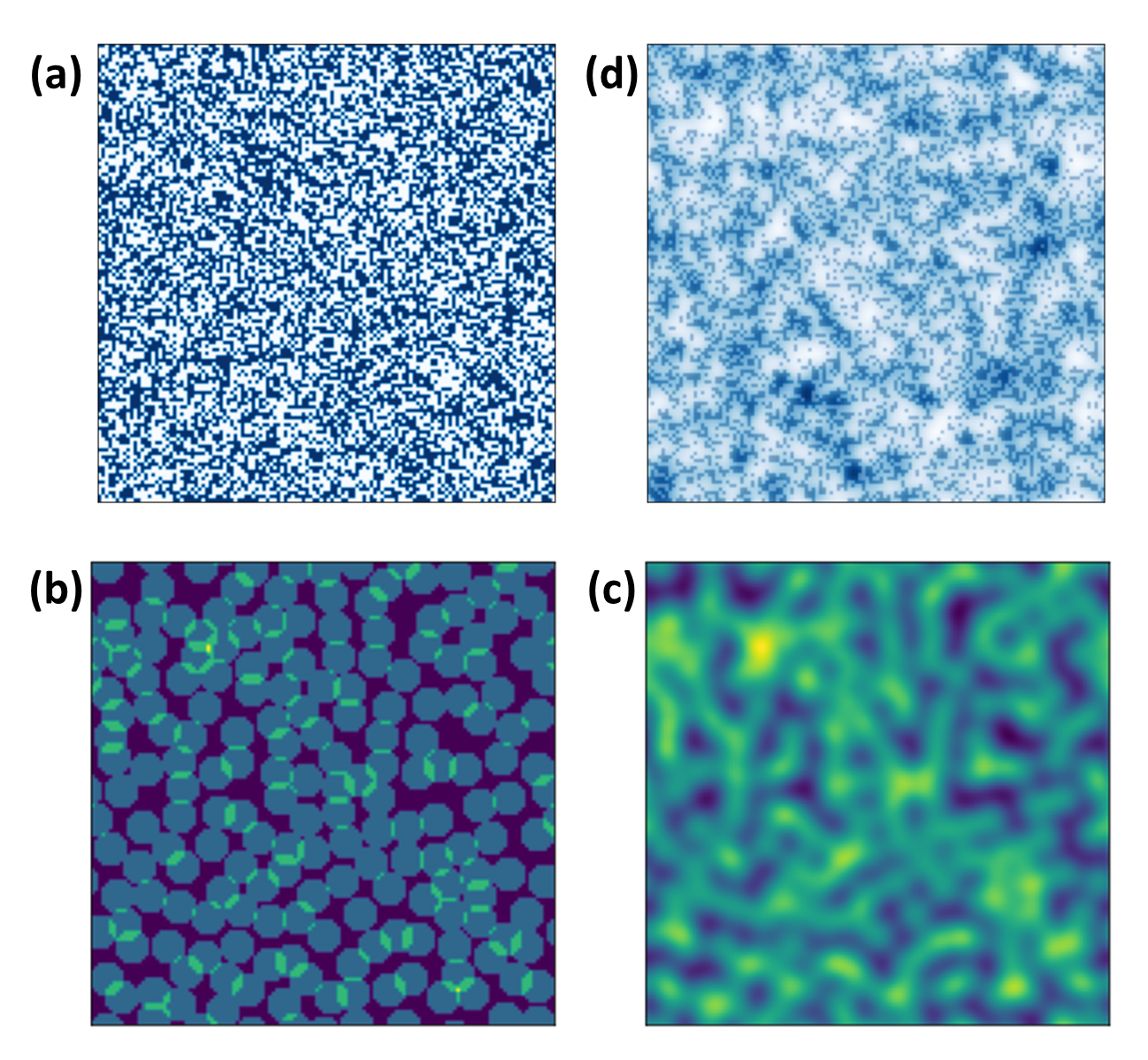}
	\caption{\emph{Perturbed Ising model reproducing experimental observations} (a) Initial domain configuration: the blue and white domains are randomly distributed and show short range density fluctuations. (b) Example of initial circle distribution for a radius of $5$ pixels, and (c) the resulting potential generated from this configuration. Yellow represents high values and blue low values. (d) When this potential is added to the Ising Hamiltonian, stronger domain density fluctuations are observed, in particular after blurring with a $2$ pixel Gaussian kernel (transparent overlay). Compare also with Fig.~4 (a) of the main text.}
	\label{SimulationSupp}
\end{figure}

To further investigate a possible origin of these density variations, we turned to simulations using the Ising model with a simple added random potential acting at a characteristic length-scale $\xi$. The parameters of the unperturbed model were chosen such that up and down polarisation regions (in equal proportion) were randomly distributed (see Fig.~\ref{SimulationSupp}(a)). The random potential was then generated as overlapping circles of specific target radius $ r = 1/2\xi$ (limiting the overlap to 25 \% of each circle's area) favouring one polarisation orientation over the other by associating a slight positive or negative potential energy with each circle (see Fig.~\ref{SimulationSupp}(b)). The circles were then blurred using a Gaussian filter to mimic a random disorder pinning potential with a characteristic correlation length $\xi$ beyond which the effect of the disorder decays (see Fig.~\ref{SimulationSupp}(c)). The resulting potential was then added to the Ising Hamiltonian, and the simulation allowed to relax (as shown in the main paper Fig.~5(a)). For each choice of random potential lengthscale $\xi$, one hundred potential configurations were generated and run.

 The resulting domain structures were blurred with a Gaussian kernel to reveal the superstructure in domain density variations (see Fig.~\ref{SimulationSupp}(d)), and the pseudo-period and typical feature size extracted from radially averaged autocorrelation analysis. As can be see in the  main paper Fig.~5(b), the feature size, given by the first minimum in the radially averaged autocorrelation function, exactly matches the characteristic lengthscale $\xi$ of the disorder potential. The characteristic lengthscale of 800 nm observed in SHG and in the domain density variations of PFM may thus provide information about the typical correlation lengths of the underlying disorder potential in the sample.

\begin{figure}[htbp]
	\centering
	\includegraphics[width=\textwidth]{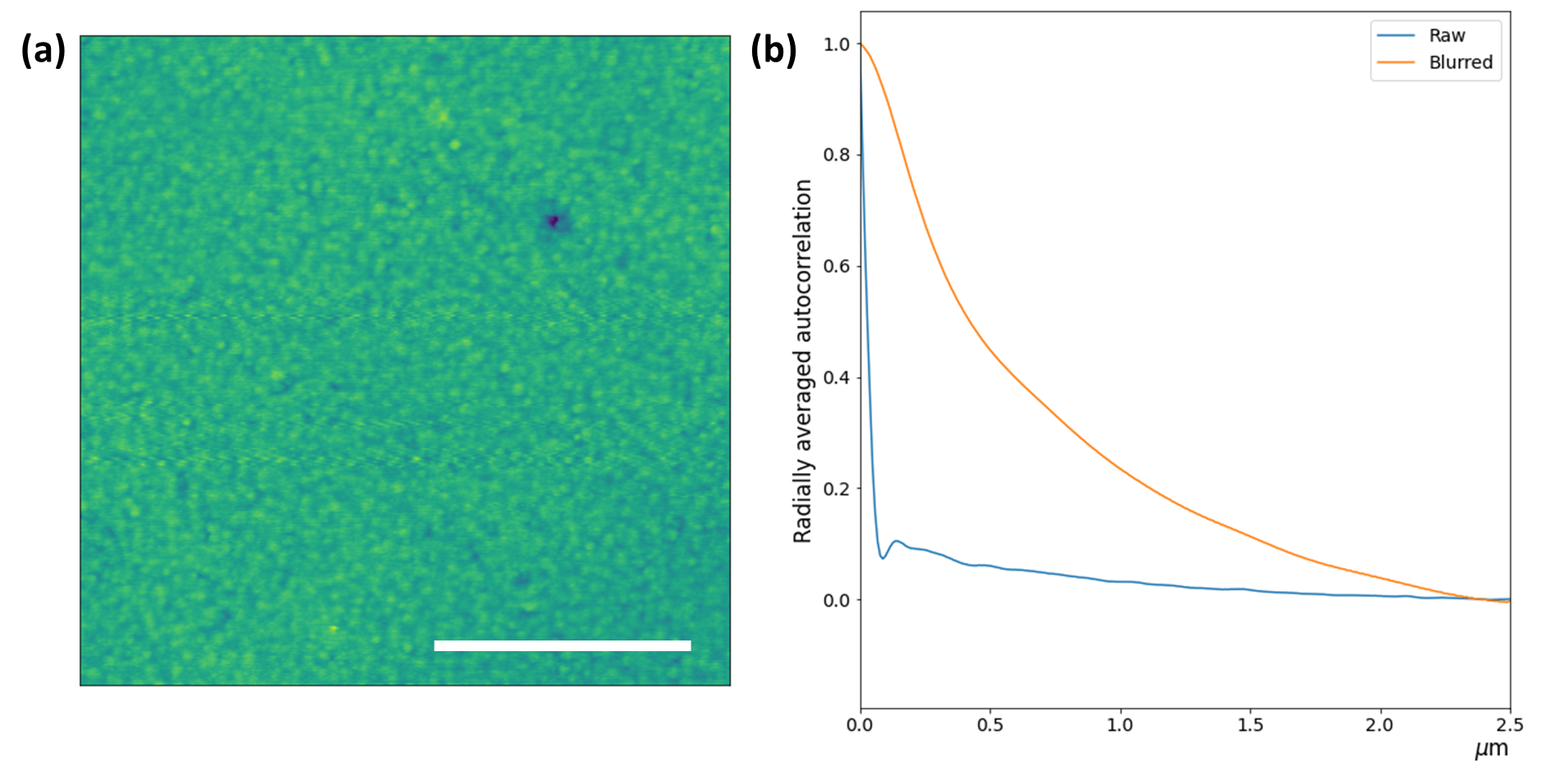}
	\caption{\emph{Surface morphology variations do not appear to influence the domain density variations} (a) Topography image acquired at the same time as the PFM in Fig.~4(a). The white scale bar represents \SI{2}{\micro\meter}. (b) Radially averaged autocorrelation of (a) for both the raw image, and the image blurred to detect density variations. The raw image shows a pseudo-period of approximately \SI{140}{\nano\meter}, but no superstructure is present in the density.}
	\label{TopoSupp}
\end{figure}

One potential source of these variations that can be readily ruled out is that the domains are pinned by the small, granular variation of surface morphology. Indeed, analysing the topography image acquired at the same time as the PFM image in Fig.~4(a) (\fref{TopoSupp}(a)) using radially averaged autocorrelation, we can see that, although the surface presents some regular features, their pseudo-period is much smaller than that of the density variations observed in the domains (\fref{TopoSupp}(b)). Furthermore, if we instead analyse the density of these topographical features by blurring them out, no structure emerges, confirming that the length scale of domain density variations is not correlated with the surface topography.

\section{Non-Ising structure and conductive properties of ferroelectric domain walls}

The specific nature of domain walls in PbTiO$_3$ is a hitherto unanswered question. Different calculations based on both ab-initio and Ginzburg-Landau-Devonshire mean field approaches have predicted various structures. The earliest calculations suggested a mixed Ising-Néel character~\cite{Lee-PRB-2009}, while more recently a room temperature Ising structure with a transition to Bloch domain walls at low temperature was predicted \cite{Wojdel-PRL-2014,Stepkova-JPCM-2012}. The question of domain wall structure in PbTiO$_3$ is an important one, as the recently proposed Bloch nature is a crucial part of the microscopic mechanism of the formation of ferroelectric skyrmions observed in this material~\cite{Das-N-2019,Goncalves-SA-2019}. 

To complicate things further, it appears that polarisation in PbTiO$_3$ can readily present complex, often rotational textures when subjected to appropriate electrostatic and strain boundary conditions~\cite{Yadav-N-2016,Hadjimichael-NatMat-2021,Hong-NanoLett-2021}. Moreover, calculations of flexoelectric effects at domain walls suggest that Néel-like polarisation discontinuities and even direct head-to-head or tail-to-tail polarisation components can be easily stabilised, in particular when inclined or strongly curved domain walls are considered~\cite{Morozovska-Ferro-2012,Eliseev-PRB-2012-85-045312,Cao-APL-2017,Feigl-NatCom-2014,Stolichnov-NanoLetters-2015}. 

We note that in the very closely related material PZT, of which PbTiO$_3$ is one of the parent compounds, there are observations of the Néel character of domain walls~\cite{Cherifi-Hertel-NC-2017,DeLuca-AdvancedMaterials-2017}, and measurements of their electrical conduction~\cite{Guyonnet-AdvMat-2011,Maksymovych-NanoLetters-2012}. Theoretical calculations suggest a strong link between the two, with (in particular charged) defects preferentially segregating at domain walls to screen the polarisation discontinuity, and providing states in the band gap to allow electric conduction. Experimentally, the currents at the PZT domain walls, as well as at domain walls in BiFeO$_3$, have been observed to increase significantly with increased defect presence, in particular via the modulation of oxygen vacancy density~\cite{Gaponenko-APL-2015,Farokhipoor-JAP-2012,Seidel-PRL-2010}.
